\documentclass[10pt,journal,doublecolumn]{IEEEtran}
\usepackage{cite}
\usepackage{amsmath,amssymb,placeins}
\usepackage{algorithmic}
\usepackage{graphicx}
\usepackage{textcomp}
\usepackage[mathscr]{euscript}
\usepackage{lipsum}
\usepackage{cuted}
\usepackage{bm}
\usepackage{multicol}
\usepackage{array}
\usepackage{makecell}
\usepackage{diagbox}
\usepackage[symbol]{footmisc}
\usepackage{tabularx}

\usepackage[mathscr]{euscript}
\usepackage[usenames]{color}
\usepackage{amsfonts}
\usepackage{latexsym}
\usepackage{subfigure}
\usepackage{multirow}
\usepackage{epstopdf}

\newtheorem{theorem}{{{\textit{Theorem}}}}
\newtheorem{note}{{{\textit{Note}}}}
\newtheorem{lemma}{{{\textit{Lemma}}}}
\newtheorem{corollary}{{{{\textit{Corollary}}}}}

\newtheorem{definition}{{{\textit{Definition}}}}

\newtheorem{remark}{{{\textit{Remark}}}}

\newtheorem{example}{{{\textit{Example}}}}

\def\BibTeX{{\rm B\kern-.05em{\sc i\kern-.025em b}\kern-.08em
		T\kern-.1667em\lower.7ex\hbox{E}\kern-.125emX}}
		
		\def\@fnsymbol#1{\ensuremath{\ifcase#1\or *\or \dagger\or \ddagger\or
   \mathsection\or \mathparagraph\or \|\or **\or \dagger\dagger
   \or \ddagger\ddagger \else\@ctrerr\fi}}

\begin{document}
\title{A Direct  and Generalized Construction of Polyphase Complementary Sets with Low PMEPR and High Code-Rate  for OFDM System}
\author{Palash~Sarkar,~
        Sudhan~Majhi,~
        and~Zilong~Liu
\thanks{Palash Sarkar is with Department of Mathematics and Sudhan Majhi is with the Department of Electrical Engineering, Indian Institute of Technology Patna, India, e-mail: {\tt palash.pma15@iitp.ac.in; smajhi@iitp.ac.in}.}
\thanks{Zilong Liu is with the School of Computer Science and Electronics Engineering, University of Essex, UK, e-mail:{\tt zilong.liu@essex.ac.uk}.}}
\IEEEpeerreviewmaketitle
\maketitle
\begin{abstract}
A major drawback of orthogonal frequency division multiplexing (OFDM) systems is their high peak-to-mean envelope power ratio (PMEPR). The PMEPR
problem can be solved by adopting large codebooks consisting of complementary sequences with low PMEPR.
In this paper, we present a new construction of polyphase complementary sets (CSs) using
generalized Boolean functions (GBFs), which generalizes Schmidt's construction in $2007$, Paterson's construction in $2000$ and Golay
complementary pairs (GCPs) given by Davis and Jedwab in $1999$. Compared with Schmidt's approach, our proposed CSs lead to lower PMEPR with higher code-rate
for sequences constructed from higher-order ($\geq 3$) GBFs. We obtain polyphase complementary sequences with maximum PMEPR of $2^{k+1}$ and
$2^{k+2}-2M$ where $k,M$ are non-negative integers that can be easily derived from the GBF associated with the CS.
\end{abstract}

\section{Introduction}
\label{sec:intro}
Orthogonal frequency-division  multiplexing (OFDM) is a
multicarrier technique which has been widely used in many high rate wireless communication standards such as Wireless Fidelity (Wi-Fi), Mobile Broadband Wireless Access (MBWA),
Worldwide Interoperability for Microwave Access (WiMax), terrestrial digital TV systems, 3GPP Long Term Evolution (LTE), etc. A major drawback of OFDM
is its large peak-to-mean envelope power ratio (PMEPR)
for the uncoded signals. PMEPR reduction through a coding perspective can be achieved by designing a large codebook whose codewords, e.g., in
the form of sequences, have low PMEPR values. In practice, OFDM signals with lower PMEPRs lead to smaller input back-off (IBO) of the power amplifier (PA) at the RF end,
thus yielding higher transmit power efficiency and larger communication range.
This paper aims to reduce PMEPR via codebooks consisting of complementary sequences which will be introduced in the sequel.

Golay complementary pair (GCP), introduced by M. J. E. Golay in \cite{golay1961}, refers to a pair of sequences whose aperiodic autocorrelation functions (AACFs) diminish to zero at each non-zero time-shift when they are summed. Either sequence
from a GCP is called a Golay sequence. The idea
of GCP was extended to complementary sets (CSs) by Tseng and Liu in \cite{chinchong} where each CS consisting of two or more constituent
sequences, called complementary sequences. A PMEPR reduction method was introduced by Davis
and Jedwab in \cite{Davis1999} to construct  standard $2^h$-ary ($h$ is a positive integer)
Golay sequences of length $2^m$ ($m$ is a positive integer) using second-order generalized Boolean function (GBF), comprising second-order cosets of
generalized first-order Reed-Muller (RM) codes $RM_{2^h}(1,m)$. By applying the constructed Golay sequences to encode  OFDM signals with
a PMEPR of at most 2, Davis and Jedwab obtained $\frac{m!}{2}2^{h(m+1)}$ codewords, called Golay-Davis-Jedwab (GDJ) code in this paper, for the phase shift keying (PSK) modulated
OFDM signals with good error-correcting capabilities, efficient encoding and decoding. Subsequently, Paterson
employed complementary sequences to enlarge the code-rate by relaxing the  PMEPR  of OFDM signal in \cite{pater2000}. Specifically, Paterson showed
that each coset of $RM_q(1,m)$ inside $RM_q(2,m)$ ($q$ is an even number no less than $2$) can be partitioned into CSs of size $2^{k+1}$ (where $k$ is a
non-negative integer depending only on $G(Q)$, a graph naturally associated with the quadratic form $Q$ in $m$ variables which defines the coset) and provided an upper bound on the PMEPR of arbitrary second-order
cosets of $RM_q(1,m)$. The construction given in \cite[Th. 12]{pater2000}\footnote{Full statement of \cite[Th. 12]{pater2000} is given in \textit{Lemma} \ref{lemmad}.} was unable to provide a tight PMEPR
bound for all the cases. By giving an improved version of \cite[Th. 12]{pater2000} in \cite[Th. 24]{pater2000}\footnote{Full statement of \cite[Th. 24]{pater2000} is given in \textit{Lemma} \ref{lemmae}.},  Paterson left the following question: \newline \textit{``What is the strongest
possible generalization of \cite[Th. 12]{pater2000}?''.}

In \cite[Th. 24]{pater2000}, it was shown that after
deleting $k$ vertices in $G(Q)$, if the resulting graph contains a path and one isolated vertex, then $Q+RM_q(1,m)$ can be partitioned into
CSs of size $2^{k+1}$ instead of $2^{k+2}$, i.e., there is no need to delete the isolated vertex.
Later, a generalization of \cite[Th. 12]{pater2000} was made by Schmidt in \cite{Schmid2007} to
establish a construction of complementary sequences that are contained in higher-order generalized RM codes. Schmidt showed in \cite{Schmid2007} that a GBF gives rise to
a CS of a given size if the graphs of all \textit{restricted Boolean functions}\footnote{A restricted Boolean function of a GBF is obtained by fixing some variables
of the GBF to some constants. If we restrict a GBF of $m$ variables over $k$ ($k<m$) fixed variables,
the restriction can be done in $2^k$ ways. Corresponding to the $2^k$ restricted Boolean functions, there are $2^k$ graphs if the restricted Boolean functions are of order $2$.}  of the GBF are paths.
In Schmidt's construction, however,
CS cannot be generated corresponding to a GBF if there is at least one restricted Boolean function whose graph is not a path (among all the restricted Boolean functions of the GBF). In this case, further restrictions need to be carried out until the graphs of all restricted Boolean functions become path.
As a result, the CS set size increases and so does the PMEPR. Because of this, a reasonable number of sequences were excluded from the Schmidt's coding scheme.
Hence, an improved version of \cite[Th. 5]{Schmid2007}\footnote{Full statement of \cite[Th. 5]{Schmid2007} is given in \textit{Lemma} \ref{lemmaf}.} or a more generalized version of \cite[Th. 12]{pater2000}
is expected to extend the range of coding options with good PMEPR bound for practical applications of OFDM.

More constructions of CSs with low PMEPR have been proposed in the literature.
In \cite{fiedler2008}, a framework has been presented to identify known Golay sequences and pairs of length $2^m$ ($m>4$) over $\mathbb{Z}_{2^h}$ in explicit algebraic normal form.
\cite{tarokh} presents a lower bound on the
PMEPR of a constant energy code as a function of its rate, distance,
and length. The results in \cite{fiedler2008} and \cite{tarokh} provide better upper bound of PMEPR than the results in \cite{pater2000}
and \cite{Schmid2007}. For multi-carrier code division multiple access (MC-CDMA), Liu \emph{et al} presented in \cite{liumc} a new class of mutually orthogonal CSs whose column sequences have PMEPR of at most 2, when each CS is arranged to be a two dimensional matrix (called a complementary matrix) whose rows constitute all of its complementary sequences in order. The low PMEPR property in Liu's construction is achieved by designing CSs such that every column sequence of a complementary matrix is a Golay sequence. Nowadays, besides polyphase complementary sequences, the design of quadrature amplitude modulation (QAM)
complementary sequences with low PMEPR is also an
interesting research topic. In \cite{imli}, QAM Golay sequences were introduced based on quadrature phase shift keying (PSK) GDJ-code. Later,
Liu \emph{et~al} constructed QAM Golay sequences by using properly selected Gaussian integer pairs \cite{liug}. Recently, numerous constructions of
complementary and quasi-/near-complementary sequences have been reported in \cite{liug,avi_commu,sdas,Sdas_lett,pskaccess,psktcom,chenzcom,chentcom,chentit,chencommlett,chenconf,ara_bzcp_2018,liuqcss,liano,liqcss1,liqcss2}.
These sequences may also be applicable
in OFDM systems to deal with the PMEPR problem, in addition to their applications in scenarios such as asynchronous communications and channel estimation.

In this paper, we propose a construction to generate new polyphase CSs with low PMEPR and high code-rate for OFDM systems by allowing both path and isolated vertices in the graphs of certain restricted versions of
higher order GBFs.
In our
proposed construction, we restrict a few number of vertices to obtain tighter PMPER. For example,  we obtain CS with maximum PMEPR of $2^{k+1}$
and $2^{k+2}-2M$ in the presence of isolated vertices  whereas the PMEPR upper bound obtained from Schmidt's construction for the same sequences
is at least $2^{k+p+1}$ (where $p$ is the number of isolated vertices present in the graphs of certain restricted Boolean functions).
The introduction of ``isolated vertices" is essential as it gives rise to higher sequence design flexibility and hence more complementary
sequences for larger code-rate, as compared to Schmidt's construction.
By moving to higher order RM code, we not only provide a partial answer to the aforementioned question raised by Paterson,
but also  extend the range of coding options for practical applications of OFDM.
It is shown that our proposed construction includes Schmidt's construction, Paterson's construction, and the GDJ code construction as special cases.
Part of this work has been presented in 2019 IEEE International Symposium on Information Theory \cite{palconf}\footnote{
In \cite{palconf}, we have presented \textit{Theorem} 1 and some preliminary results derived from it. Based on \cite{palconf}, we further provide a graphical analysis of our proposed construction.
Moreover, we construct codes
with maximum PMEPR $4$, $6$, and $8$, and compare the proposed code-rates with the existing constructions \cite{pater2000} and \cite{Schmid2007}.}.

The remainder of the paper is organized as follows. In Section II, some useful notations
and definitions are given. In Section III, a generalized construction of CS is presented. Section IV contains some results which are obtained from our proposed construction.
We have presented a graphical analysis of our proposed construction in Section V.
Then we compare our proposed construction with \cite{pater2000,Schmid2007} in Section VI. Finally,
concluding remarks are drawn in Section VII.
\section{Preliminary}
\subsection{Notations}
The following notations will be used throughout this paper:
\begin{itemize}
 \item $J=\{j_0,j_1,\hdots,j_{k-1}\}\subset\{0,1,\hdots,m-1\}$.
 \item $\textbf{x}_J=(x_{j_0},x_{j_1},\hdots,x_{j_{k-1}})$.
 \item $\textbf{c}=(c_0,c_1,\hdots,c_{k-1})\in \{0,1\}^k$.
 \item $\textbf{d}=(d_0,d_1,\hdots,d_{k-1})\in\{0,1\}^k$.
 \item $\omega_q=\exp(2\pi \sqrt{-1}/q)$, $q\geq 2, 2|q $.
\end{itemize}

\label{sec:back}
\subsection{Definitions of Correlations and Sequences}
Let $\textbf{a}=(a_0,a_1,\hdots, a_{L-1})$ and $\textbf{b}=(b_0,b_1,\hdots,b_{L-1})$ be two complex-valued sequences of equal length $L$ and
let $\tau$ be an integer. Define
\begin{equation}\label{equ:cross}
\begin{split}
C(\textbf{a}, \textbf{b})(\tau)=\begin{cases}
\sum_{i=0}^{L-1-\tau}a_{i+\tau}b^{*}_i, & 0 \leq \tau < L, \\
\sum_{i=0}^{L+\tau -1} a_ib^{*}_{i-\tau}, & -L< \tau < 0,  \\
0, & \textnormal{otherwise},
\end{cases}
\end{split}
\end{equation}
and $A(\textbf{a})(\tau)=C(\textbf{a},\textbf{a})(\tau)$.
The above mentioned functions are called the aperiodic cross-correlation function
between $\textbf{a}$ and $\textbf{b}$ and the AACF of $\textbf{a}$, respectively.
\begin{definition}
A set of $n$ sequences $\textbf{a}^0,\textbf{a}^1, \hdots ,\textbf{a}^{n-1}$, each of equal length $L$, is said to be a CS if
\begin{equation}
 \begin{split}
  A(\textbf{a}^0)(\tau)\!+\!A(\textbf{a}^1)(\tau)\!+\!\hdots \!+\!A(\textbf{a}^{n-1})(\tau)\!\!=\!\!
  \begin{cases}
   nL, & \tau=0,\\
   0, & \textnormal{otherwise}.
  \end{cases}
 \end{split}
\end{equation}
\end{definition}
A CS of size two is called a GCP.
\subsection{PMEPR of OFDM signal}
For $q$-PSK modulation, the OFDM signal for the word $\textbf{a}=(a_0,a_1,\hdots, a_{L-1})$ (where $a_i\in \mathbb{Z}_q$) can be modeled as the real part
of
\begin{equation}\nonumber
S(\textbf{a})(t)=\displaystyle\sum_{\alpha=0}^{L-1}\omega_q^{a_{\alpha}} \exp\left(2\pi \sqrt{-1}(f_0+\alpha f_s)t\right),
\end{equation}
where $0\leq t<T$ (where $T$ denotes the OFDM symbol duration), $f_0$ denotes the center carrier frequency, and $f_s$ the subcarrier spacing.
We define the instantaneous envelope power of the OFDM signal as \cite{pater2000}
 \begin{equation} \nonumber
  P(\textbf{a})(t)=|S(\textbf{a})(t)|^2.
 \end{equation}
 From the above expression, it is easy to derive that
 \begin{equation}\label{pmeprdef}
 \begin{split}
  P(\textbf{a})(t)=&\displaystyle\sum_{\tau=1-L}^{L-1}A(\textbf{a})(\tau)\exp(2\pi \sqrt{-1}\tau f_s t)\\
  =&A(\textbf{a})(0)\!\!+\!\!2\!\cdot\! \text{Re}\left\{ \displaystyle\sum_{\tau=1}^{L-1}\!\!\!A(\textbf{a})(\tau)\exp(2\pi \sqrt{-1}\tau f_s t)\right\},
\end{split}
\end{equation}
where $\text{Re}\{x\}$ denotes the real part of a complex
number $x$. We define the PMEPR of the signal
$S(\textbf{a})(t)$ as
\begin{equation}\label{defipmeprnew}
\textnormal{PMEPR}(\textbf{a})=\frac{1}{L}\displaystyle \sup_{0\leq f_s t<1}P(\textbf{a})(t).
\end{equation}
The peak amplitude of an $L$-subcarrier OFDM signal is $L$.
\subsection{Generalized Boolean Functions}
Let $f$ be a function of $m$ variables $x_0,x_1,\hdots,x_{m-1}$ over $\mathbb{Z}_q$. A monomial of degree $r$ is defined as the product
of any $r$ distinct variables among  $x_0,x_1,\hdots, x_{m-1}$. There are $2^m$ distinct monomials over $m$ variables listed below:
$1,x_0,x_1,\hdots,x_{m-1},x_0x_1,x_0x_2,\hdots,x_{m-2}x_{m-1},\hdots,\\x_0x_1\hdots x_{m-1}.$
A function $f$ is said to be a GBF of order $r$ if it can be uniquely expressed as a linear
combination of monomials of degree at most $r$, where the coefficient of each monomial is drawn from $\mathbb{Z}_q$. A GBF of order $r$ can be
expressed as
\begin{equation}\label{rorder}
 f=Q+\sum_{i=0}^{m-1}g_ix_i+g',
\end{equation}
where
\begin{equation}
 \begin{split}
  Q\!=\!\sum_{p=2}^{r}\sum_{0\leq \alpha_0\!<\!\alpha_1\!<\!\hdots\!<\!\alpha_{p-1}\!<\!m}\!\!\!\!\!\!\!\!\!\!\!\!\!\!\!\!a_{\alpha_0,\alpha_1,\hdots,\alpha_{p-1}}x_{\alpha_0}x_{\alpha_1}\hdots x_{\alpha_{p-1}},
 \end{split}
\end{equation}
and $g_i,g',a_{\alpha_0,\alpha_1,\hdots,\alpha_{p-1}}\in \mathbb{Z}_q$.
\subsection{Quadratic Forms and Graphs}
Let $f$ be a $r$th order GBF of $m$ variables over $\mathbb{Z}_q$.
Then $f\big\arrowvert_{\textbf{x}_J=\textbf{c}}$ is obtained by substituting $x_{j_\alpha}=c_\alpha$ ($\alpha=0,1,\hdots,k-1$) in $f$.
If $f\big\arrowvert_{\textbf{x}_J=\textbf{c}}$ is a quadratic GBF, then $G(f\big\arrowvert_{\textbf{x}_J=\textbf{c}})$ denotes a graph with
$V=\{x_0,x_1,\hdots,x_{m-1}\}\setminus \{x_{j_0},x_{j_1},\hdots,x_{j_{k-1}}\}$ as the set of vertices. The $G(f\big\arrowvert_{\textbf{x}_J=\textbf{c}})$
is obtained by joining the vertices $x_{\alpha_1}$ and $x_{\alpha_2}$ by an edge if
there is a term $q_{\alpha_1\alpha_2}x_{\alpha_1}x_{\alpha_2}$ ($0\leq \alpha_1<\alpha_2<m$, $x_{\alpha_1}$, $x_{\alpha_2}\in V$) in the
GBF $f\big\arrowvert_{\textbf{x}_J=\textbf{c}}$ with $q_{\alpha_1\alpha_2}\neq 0$ ($q_{\alpha_1\alpha_2}\in \mathbb{Z}_q$). For $k=0$, $G(f\big\arrowvert_{\textbf{x}_J=\textbf{c}})$
is nothing but $G(f)$.
\subsection{Sequence Corresponding to a Generalized Boolean Function}
Corresponding to a GBF $f$,
we define a complex-valued vector (or sequence) $\psi(f)$, as follows.
\begin{equation}
 \psi(f)=(\omega_q^{f_0}, \omega_q^{f_1},\hdots, \omega_q^{f_{2^m-1}}),
\end{equation}
where $f_i=f(i_0,i_1,\hdots,i_{m-1})$ and $(i_0,i_1,\hdots,i_{m-1})$ is the binary vector representation of integer
$i$ ($i=\sum_{\alpha=0}^{m-1}i_\alpha 2^\alpha$). Throughout the paper, even $q$ not less than $2$ will be considered.

Again, we define $\psi(f\big\arrowvert_{\textbf{x}_J=\textbf{c}})$ as a complex-valued sequence with $\omega_q^{f(i_0,i_1,\hdots,i_{m-1})}$ as $i$th
component if $i_{j_\alpha}=c_\alpha$ for each $0\leq \alpha <k$ and equal to zero otherwise.
\begin{definition}[Effective-Degree of a GBF \cite{Schmid2007}]
The effective-degree of a GBF $f:\{0,1\}^m\rightarrow \mathbb{Z}_{2^h}$, is defined as follows.
\begin{equation}
 \max_{0\leq i<h}[\deg\left(f\!\!\!\!\!\mod 2^{i+1}\right)-i].
\end{equation}
\end{definition}
Let $\mathcal{F}(r,m,h)$ be the set of all GBFs $f:\{0,1\}^m\rightarrow \mathbb{Z}_{2^h}$. Also, let $|\mathcal{F}(r,m,h)|$ denote
the number of GBFs in $\mathcal{F}(r,m,h)$ which is given by \cite{Schmid2007}
\begin{equation}
 \log_2|\mathcal{F}(r,m,h)|=\sum_{i=0}^r h\binom{m}{i}+\sum_{i=1}^{h-1}(h-i)\binom{m}{r+i}.
\end{equation}
\begin{definition}[Effective-Degree RM Code \cite{Schmid2007}]
 For $0\leq r\leq m$, the effective-degree RM code is denoted by ERM$(r,m,h)$ and defined as
 \begin{equation}
  \textnormal{ERM}(r,m,h)=\{\psi(f): f\in \mathcal{F}(r,m,h)\}.
 \end{equation}
\end{definition}
\begin{definition}[Lee Weight and Squared Euclidean Weight]
 Let $\textbf{a}=(a_0,a_1,\hdots, a_{L-1})$ be a $\mathbb{Z}_{2^h}$-valued sequence. The Lee weight of $\textbf{a}$ is denoted by $wt_L(\textbf{a})$ and
 defined as follows.
 \begin{equation}
  wt_L(\textbf{a})=\sum_{i=0}^{L-1}\min\{a_i,2^h-a_i\}.
 \end{equation}
\end{definition}
The squared Euclidean weight of $\textbf{a}$ (when the entries of $\textbf{a}$ are mapped onto a $2^h$-ary PSK constellation) is denoted by $wt^2_E(\textbf{a})$
and given by
\begin{equation}
 wt^2_E(\textbf{a})=\sum_{i=0}^{L-1}|\omega_q^{a_i}-1|^2.
\end{equation}
Let $d_L(\textbf{a},\textbf{b})=wt_L(\textbf{a}-\textbf{b})$ and $d^2_E(\textbf{a},\textbf{b})=wt^2_E(\textbf{a}-\textbf{b})$ be the Lee and squared Euclidean
distance between $\textbf{a},\textbf{b}\in \mathbb{Z}^L_{2^h}$, respectively. The symbols $d_L(\mathcal{C})$ and $d^2_E(\mathcal{C})$ will be used to denote
minimum distances (taken over all distinct sequences) of a code $\mathcal{C}\in \mathbb{Z}^L_{2^h}$.

Next, we present some lemmas which will be used in our proposed construction.
\begin{lemma}[\cite{pater2000}]\label{lemmac}
Let $f,g$ be two GBFs of $m$ variables.
Consider $0\leq i_0<i_1<\cdots< i_{l-1}<m$, which is a list of $l$ indices and the set $\{i_0,i_1,\hdots,i_{l-1}\}$ has no intersection with  $J$.
Let $\textbf{y}=(x_{i_0},x_{i_1},\hdots,x_{i_{l-1}})$, then
\begin{equation}
\begin{split}
C&\left (\psi(f\arrowvert_{\textbf{x}_J=\textbf{c}}),\psi(g\arrowvert_{\textbf{x}_J=\textbf{d}})\right )(\tau)\\&=\displaystyle\sum_{\textbf{c}_1,\textbf{c}_2}
C\left(\psi(f\arrowvert_{\textbf{xy}=\textbf{cc}_1}),\psi(g\arrowvert_{\textbf{xy}=\textbf{dc}_2})\right)(\tau).
\end{split}
\end{equation}
\end{lemma}
\begin{lemma}[\cite{stinch}] \label{lemmaa}
Suppose that there are two GBFs $f$ and $f'$ of $m$-variables $x_0,x_1,\hdots,x_{m-1}$ over $\mathbb{Z}_q$,
such that for $k\leq m-3$, $f\big\arrowvert _{\textbf{x}_J=\textbf{c}}$
and $f'\big\arrowvert_{\textbf{x}_J=\textbf{c}}$ are given by
\begin{equation}
\begin{split}
f\big\arrowvert _{\textbf{x}_J=\textbf{c}}&=P+L+g_lx_l+g,\\
f'\big\arrowvert _{\textbf{x}_J=\textbf{c}}&=P+L+g_lx_l+\frac{q}{2}x_a+g,
\end{split}
\end{equation}
where $L=\sum_{\alpha =0}^{m-k-2}g_{i_\alpha} x_{i_\alpha}$, $\{i_0,i_1,\cdots,i_{m-k-2}\}\!\!\!=\{0,1,\hdots,m-1\}\setminus J\cup\{l\}$, both  $G(f\big\arrowvert _{\textbf{x}_J=\textbf{c}})$ and $G(f'\big\arrowvert _{\textbf{x}_J=\textbf{c}})$
consist of a path over $m-k-1$ vertices, given by $G(P)$, $x_a$ is an either end vertex, $x_l$ is an isolated vertex,
and $g_l, g\in \mathbb{Z}_q$. Then for fixed $\textbf{c}$ and $d_1\neq d_2$ $(d_1, d_2\in\{0,1\})$,
\begin{equation}
\begin{split}
C&\!(f\big\arrowvert _{\textbf{x}_Jx_l=\textbf{c}d_1}\!,\!f\big\arrowvert _{\textbf{x}_Jx_l=\textbf{c}d_2})\!(\tau)\!\!+\!\!
C\!(f'\big\arrowvert _{\textbf{x}_Jx_l=\textbf{c}d_1}\!,\!f'\big\arrowvert _{\textbf{x}_Jx_l=\textbf{c}d_2})\!(\tau)
\\&=
\begin{cases}
\omega_q^{(d_1-d_2)g_l}2^{m-k}, & \tau=(d_2-d_1)2^l,\\
0, & \textnormal{otherwise}.
\end{cases}
\end{split}
\end{equation}
\end{lemma}
\begin{lemma}[\cite{rati}]\label{lemmab}
Let $\textbf{c}_1,\textbf{c}_2$ $\in \{0,1\}^k$. If $\textbf{c}_1\neq \textbf{c}_2$,
$\displaystyle\sum_{\textbf{d}}(-1)^{\textbf{d}\cdot(\textbf{c}_1+\textbf{c}_2)}=0$.
\end{lemma}
\begin{lemma}[\cite{Davis1999}]
Suppose that $f:\{0,1\}^m\rightarrow \mathbb{Z}_q$ is a quadratic GBF of $m$ variables. Suppose further that $G(f)$ is a path with $2^{h-1}$
being the weight of every edge. Then for any choice of $c,c'\in$ $\mathbb{Z}_{2^h}$, the pair $$\left(f+c,f+2^{h-1}x_a+c'\right)$$ forms a GCP.
\end{lemma}
\begin{lemma}[{\cite[Th. 12]{pater2000}}]\label{lemmad}
 Suppose that $f:\{0,1\}^m\rightarrow \mathbb{Z}_q$ is a quadratic GBF of $m$ variables. Suppose further that $G(f)$ contains a set of $k$ distinct
 vertices labeled $j_0, j_1, \hdots, j_{k-1}$ with the property that deleting those $k$ vertices and corresponding their edges results in a path.
 Then for any choice of $g_i$, $g'\in $ $\mathbb{Z}_q$, where $g_i$ is the coefficient of $x_i$ and $g'$  is a constant
 term in $f$, we have
\begin{equation}
\left\{f+\frac{q}{2}\left(\sum_{\alpha=0}^{k-1}d_{\alpha}x_{j_{\alpha}}+d''x_a    \right): d_{\alpha}, d''\in \{0,1\} \right\}
\end{equation}
is a CS of size $2^{k+1}$.
\end{lemma}
\begin{lemma}[{\cite[Th. 24]{pater2000}}]\label{lemmae}
Suppose that $f:\{0,1\}^m\rightarrow \mathbb{Z}_q$ is a quadratic GBF of $m$ variables. In addition, suppose that $G(f)$ contains a set of $k$ distinct
vertices labeled $j_0, j_1, \hdots, j_{k-1}$ with the property that deleting those $k$ vertices and all their edges results in a path on $m-k-1$
vertices and an isolated vertex. Suppose further that all edges in the original graph between the isolated vertex and the $k$ deleted vertices
are weighted by $q/2$. Let $x_a$ be the either end vertex in this path. Then for any choice of $g_i$, $g'\in $ $\mathbb{Z}_q$
\begin{equation}
\left\{f+\frac{q}{2}\left(\sum_{\alpha=0}^{k-1}d_{\alpha}x_{j_{\alpha}}+d''x_a    \right): d_{\alpha}, d'' \in \{0,1\} \right\}
\end{equation}
is a CS of size $2^{k+1}$.
\end{lemma}
\begin{lemma}[{\cite[Th. 5]{Schmid2007}}]\label{lemmaf}
Let $f:\{0,1\}^m\rightarrow \mathbb{Z}_q$ be a GBF of $m$ variables. Suppose further that for each $\textbf{c}\in \{0,1\}^k$,
$G(f\big\arrowvert_{\textbf{x}_J=\textbf{c}})$ is a path in $m-k$ vertices. Suppose further that $q/2$ is the weight of each edge
of the path $G(f\big\arrowvert_{\textbf{x}_J=\textbf{c}})$. Then for any choice of $g_i$, $g'\in $ $\mathbb{Z}_q$
\begin{equation}\label{papa}
\left\{f+\frac{q}{2}\left(\sum_{\alpha=0}^{k-1}d_{\alpha}x_{j_{\alpha}}+d''e_1    \right): d_{\alpha}, d'' \in \{0,1\} \right\}
\end{equation}
is a CS of size $2^{k+1}$ and hence $\psi(f)$ lies in a CS of size $2^{k+1}$. In (\ref{papa}), $e_1$ is a function given by
\begin{equation}\nonumber
 {e_1=\displaystyle\sum_{\textbf{c}\in\{0,1\}^k}x_{\pi_{\textbf{c}}(0)}\prod_{\alpha=0}^{k-1}x_{j_\alpha}^{c_\alpha}(1-x_{j_\alpha})^{(1-c_\alpha)},}
\end{equation}
where $\pi_{\textbf{c}}$, $\textbf{c}\in\{0,1\}^k$, are $2^k$ permutaions of $\{0,1,\hdots,m-1\}\setminus J$, which may or may not be distinct.
Note that $e_1\arrowvert_{\textbf{x}_J=\textbf{c}}=x_{\pi_{\textbf{c}}(0)}$ is one of the end vertices
in the path $G(f\big\arrowvert_{\textbf{x}_J=\textbf{c}})$, where $G(f\big\arrowvert_{\textbf{x}_J=\textbf{c}})$ is identified by the quadratic form
$\left(\displaystyle\frac{q}{2}\sum_{\alpha=0}^{m-k-2}x_{\pi_{\textbf{c}}(\alpha)}x_{\pi_{\textbf{c}}(\alpha+1)}\right)$.
It is also noted that for given $\pi_\textbf{c}$, we have
$e_1\arrowvert_{\textbf{x}_J=\textbf{c}}=x_{\pi_{\textbf{c}}(m-k-1)}$ if the reversed permutation of $\pi_\textbf{c}$ is chosen.
\end{lemma}
\begin{lemma}[{\cite[Th. 9]{Schmid2007}}]
\begin{equation}
 \begin{split}
  d_L(\textnormal{ERM}(r,m,h))&=2^{m-r},\\
  d_E^2(\textnormal{ERM}(r,m,h))&=2^{m-r+2}\sin^2\left(\frac{\pi}{2^h}\right).
 \end{split}
\end{equation}
\end{lemma}

\section{Proposed Constructions}
In this section, we present a generalized construction of CS. For ease of presentation, whenever the context is clear, we
use $C(f,g)(\tau)$ to denote $C(\psi(f),\psi(g))(\tau)$ for any two GBFs $f$ and $g$. Similar changes are applied to restricted
Boolean functions as well.
\begin{theorem}\label{Theorem1}
Let $f$ be a GBF of $m$ variables over $\mathbb{Z}_q$ with the property that there exist $M$ number of such $\textbf{c}$
for which  $G(f\big\arrowvert_{\textbf{x}_J=\textbf{c}})$ is a path over $m-k$ vertices and there exist $N_i$
number of such $\textbf{c}$ for which $G(f\big\arrowvert_{\textbf{x}_J=\textbf{c}})$ consists of a path over $m-k-1$ vertices and one isolated vertex $x_{l_i}$
such that $M,N_i\geq 0, M+\displaystyle\sum_{i=1}^pN_i=2^k$.
Suppose further that all the relevant edges in $G(f\big\arrowvert_{\textbf{x}_J=\textbf{c}})$ (for all $\textbf{c}$) have identical weight of $q/2$. Then
for any choice of $g_i,g'\in \mathbb{Z}_q$, $\psi(f)$ lies in a set $S$ of size $2^{k+1}$ with the following
aperiodic auto-correlation property.
\begin{equation}\label{autocorrelationofS}
 \begin{split}
  A(S)(\tau)\!\!=\!\!\begin{cases}
        2^{m+k+1},& \tau=0,\\
        \omega_q^{g_{l_i}}2^m\displaystyle\sum_{\textbf{c}\in S_{N_i}}\omega_q^{L^{l_i}_{\textbf{c}}}, & \tau\!=\!2^{l_i},i\!\!=\!\!1,2,\hdots,p,\\
        \omega_q^{-g_{l_i}}2^m\displaystyle\sum_{\textbf{c}\in S_{N_i}}\omega_q^{-L^{l_i}_{\textbf{c}}}, & \tau=-2^{l_i},i\!\!=\!\!1,2,\!\hdots\!,p,\\
        0, & \textnormal{otherwise},
       \end{cases}
 \end{split}
\end{equation}
where $g_{l_i} \in \mathbb{Z}_q$, $i=1,2,\hdots,p$, is the coefficient of $x_{l_i}$ in $f$, $S_{N_i}$ contains all those $\textbf{c}$ for
which $G(f\big\arrowvert_{\textbf{x}_J=\textbf{c}})$ consists of a path over $m-k-1$ vertices and one isolated vertex
labeled $l_i$ ($l_i\in\{0,1,\hdots,m-1\}\setminus J$, and $l_1,l_2,\hdots,l_p$ are all distinct),
and
\begin{equation}\nonumber
\begin{split}
L^{l_i}_{\textbf{c}}\!\!=\!\!\!&\displaystyle\sum_{r=1}^k\sum_{0\leq i_1<i_2<\cdots<i_r<k}\!\!\!\!\!\!\!\!\!\!\!\!\!\!
\varrho^{l_i}_{i_1,i_2,\hdots,i_r}c_{i_1}c_{i_2}\cdots c_{i_r}\!~\!(\varrho^{l_i}_{i_1,i_2,\hdots,i_r}\textnormal{'s}\in \!\mathbb{Z}_q),
\end{split}
\end{equation}
where $L^{l_i}_{\textbf{c}}$ is obtained by setting $\textbf{x}_J=\textbf{c}$ in $L^{l_i}_{\textbf{x}_J}$ which is a function
and associated with the variables
$x_{j_0},x_{j_1},\hdots,x_{j_{k-1}}$ and $x_{l_i}$. The term $L^{l_i}_{\textbf{x}_J}$ can be expressed as
$\displaystyle\sum_{r=1}^k\sum_{0\leq i_1<i_2<\cdots<i_r<k}\!\!\!\!\!\!\!\!\!\!\!\!\!\!
\varrho^{l_i}_{i_1,i_2,\hdots,i_r}x_{i_1}x_{i_2}\cdots x_{i_r}$.
\end{theorem}
\begin{IEEEproof}
See Appendix A.
\end{IEEEproof}
We have introduced $M$ and $N_i$ ($i=1,2,\hdots,p$) in \textit{Theorem} \ref{Theorem1} with the condition $M+\displaystyle\sum_{i=1}^pN_i=2^k$, $M,N_i\geq 0$.
Therefore, $M$ and $N_i$'s range from $0$ to $2^k$.
\begin{remark}[Explicit Form of GBFs and the set $S$ as Defined in \textit{Theorem} \ref{Theorem1}]
The GBF $f$, as defined in \textit{Theorem} \ref{Theorem1}, can be expressed as
\begin{equation}\label{cgbf}
\begin{split}
&\frac{q}{2}\sum_{\textbf{c}\in S_M}\sum_{i=0}^{m-k-2}x_{\pi_{\textbf{c}}(i)}
 x_{\pi_{\textbf{c}}(i+1)}\prod_{\alpha=0}^{k-1}x_{j_\alpha}^{c_\alpha}(1-x_{j_\alpha})^{(1-c_\alpha)}\\
 &+\frac{q}{2}\sum_{\delta=1}^p\sum_{\textbf{c}\in S_{N_{\delta}}}\sum_{i=0}^{m-k-3}\!\!\!\! x_{\pi^\delta_{\textbf{c}}(i)}
 x_{\pi^\delta_{\textbf{c}}(i+1)}\prod_{\alpha=0}^{k-1}x_{j_\alpha}^{c_\alpha}(1-x_{j_\alpha})^{(1-c_\alpha)}\\&+
 \sum_{\delta=1}^p\sum_{r=1}^k\sum_{0\leq i_1<i_2<\cdots<i_r<k}\!\!\!\!\!\!\!\!\!\!\!\!\!\varrho^{l_\delta}_{i_1,i_2,\hdots,i_r}x_{j_{i_1}}x_{j_{i_2}}\cdots x_{j_{i_r}}x_{l_\delta}\\&+
 \sum_{r=2}^k\sum_{0\leq i_1<i_2<\cdots<i_r<k}\!\!\!\!\!\!\!\!\!\!\!\!\!\alpha_{i_1,i_2,\hdots,i_r}x_{j_{i_1}}x_{j_{i_2}}\cdots x_{j_{i_r}}\!\!\!\!+\!\!
\sum_{i=0}^{m-1}g_ix_i+g',
 \end{split}
\end{equation}
where $\pi^\delta_{\textbf{c}}$ are $N_\delta$ permutations of
$\{0,1,\hdots,m-1\}\setminus J\cup\{l_\delta\}$ ($\delta=1,2,\hdots,p$),
$\pi_{\textbf{c}}$ are $M$ permutations of $\{0,1,\hdots,m-1\}\setminus J$, and $\alpha_{i_1,i_2,\hdots,i_r}$'s belong to $\mathbb{Z}_q$.
The set $S$ can be expressed as
\begin{equation}\label{defis}
S=\left \{f+\frac{q}{2}\left(\mathbf{d}\cdot\mathbf{x}_J+d''e_2\right): \mathbf{d}\in \{0,1\}^k, d''\in\{0,1\}\right\},
\end{equation}
where $\mathbf{d}\cdot\mathbf{x}_J=\displaystyle{\sum_{\alpha=0}^{k-1}} d_{\alpha}x_{j_{\alpha}}$.
In (\ref{defis}), $e_2$ is the function given by
\begin{equation}\label{deter_ev}
\begin{split}
 e_2=&\displaystyle\sum_{\textbf{c}\in S_M}x_{\pi_{\textbf{c}}(0)}\prod_{\alpha=0}^{k-1}x_{j_\alpha}^{c_\alpha}(1-x_{j_\alpha})^{(1-c_\alpha)}\\
 &+\sum_{\delta=1}^p\sum_{\textbf{c}\in S_{N_{\delta}}}x_{\pi^\delta_{\textbf{c}}(0)}\prod_{\alpha=0}^{k-1}x_{j_\alpha}^{c_\alpha}(1-x_{j_\alpha})^{(1-c_\alpha)}.
\end{split}
 \end{equation}
It is to be noted that $e_2\arrowvert_{\textbf{x}_J=\textbf{c}}=x_{\pi_{\textbf{c}}(0)}$ which is one of the end vertices in the
path $G(f\arrowvert_{\textbf{x}_J=\textbf{c}})$ for $\textbf{c}\in S_M$. Similarly, $e_2\arrowvert_{\textbf{x}_J=\textbf{c}}=x_{\pi^\delta_{\textbf{c}}(0)}$
which is one of the end vertices of the path lying in $G(f\arrowvert_{\textbf{x}_J=\textbf{c}})$ for $\textbf{c}\in S_{N_{\delta}}$ ($\delta=1,2,\cdots,p$).
%
\end{remark}
From the expression of the GBF $f$ given in (\ref{cgbf}), we have the following observations:

For $\textbf{c}\in S_M$, $G(f\big\arrowvert_{\textbf{x}_J=\textbf{c}})$ is a path over $m-k$ vertices and the path is identified by
the quadratic term $\frac{q}{2}\sum_{i=0}^{m-k-2}x_{\pi_{\textbf{c}}(i)}
 x_{\pi_{\textbf{c}}(i+1)}$. As the size of $S_M$ is $M$, $\textbf{c}$ has $M$ choices in $S_M$. We assume that
 $\textbf{c}_0,\textbf{c}_1,\hdots,\textbf{c}_{M-1}$ are the $M$ choice of $\textbf{c}$ in $S_M$, i.e.,
 $S_M=\{\textbf{c}_0,\textbf{c}_1,\hdots,\textbf{c}_{M-1}\}$. For $M$ vectors in $S_M$, we get $M$ restricted Boolean functions
 $f\big\arrowvert_{\textbf{x}_J=\textbf{c}_i}$, $i=0,1,\hdots,M-1$, which may or may not be distinct and corresponding to each restricted
 Boolean function, we get a path. Therefore, the term $\frac{q}{2}\sum_{\textbf{c}\in S_M}\sum_{i=0}^{m-k-2}x_{\pi_{\textbf{c}}(i)}
 x_{\pi_{\textbf{c}}(i+1)}\prod_{\alpha=0}^{k-1}x_{j_\alpha}^{c_\alpha}(1-x_{j_\alpha})^{(1-c_\alpha)}$, present in $f$, generates the paths
 $G(f\big\arrowvert_{\textbf{x}_J=\textbf{c}})$ for $\textbf{c}\in S_M$.

Similarly, $\frac{q}{2}\sum_{\textbf{c}\in S_{N_{\delta}}}\sum_{i=0}^{m-k-3}x_{\pi^\delta_{\textbf{c}}(i)}
 x_{\pi^\delta_{\textbf{c}}(i+1)}\prod_{\alpha=0}^{k-1}x_{j_\alpha}^{c_\alpha}(1-x_{j_\alpha})^{(1-c_\alpha)}$ generates $N_\delta$ graphs,
 denoted by $G(f\big\arrowvert_{\textbf{x}_J=\textbf{c}})$, $\textbf{c}\in S_{N_\delta}$, where each of $N_\delta$ graphs contains one path
 and one isolated vertex $x_{l_\delta}$. It is noted that the paths in $N_\delta$ graphs may or may not be distinct, it depends on the
 permutations $\pi^\delta_{\textbf{c}}$, $\textbf{c}\in S_{N_\delta}$. Therefore, the term $\frac{q}{2}\sum_{\delta=1}^p\sum_{\textbf{c}\in S_{N_{\delta}}}\sum_{i=0}^{m-k-3}x_{\pi^\delta_{\textbf{c}}(i)}
 x_{\pi^\delta_{\textbf{c}}(i+1)}\prod_{\alpha=0}^{k-1}x_{j_\alpha}^{c_\alpha}(1-x_{j_\alpha})^{(1-c_\alpha)}$ generates
 $\sum_{i=1}^p N_i$ graphs, where each of $N_i$ graphs contains a path and one isolated vertex
 $x_{l_i}$, $i=1,2,\hdots,p$.

From the expression of $f$, it can easily be observed that $x_{j_0},x_{j_1},\hdots,x_{j_{k-1}}$ are the restricted variables. Below we
have listed
$2^k-1$ distinct monomials over the $k+1$ variables $x_{j_0},x_{j_1},\hdots,x_{j_{k-1}}$ and $x_{l_\delta}$:
$x_{j_0}x_{l_\delta},x_{j_1}x_{l_\delta},\hdots,x_{j_{k-1}}x_{l_\delta},x_{j_0}x_{j_1}x_{l_\delta},x_{j_0}x_{j_2}x_{l_\delta},\hdots,\\x_{j_{k-2}}x_{j_{k-1}}x_{l_\delta},\hdots,x_{j_0}x_{j_1}\cdots x_{j_{k-1}}x_{l_\delta}.$
Now, we consider the following term:
\begin{equation}\nonumber
 \begin{split}
  \sum_{r=1}^k\sum_{0\leq i_1<i_2<\cdots<i_r<k}\!\!\!\!\!\!\!\!\!\!\!\!\!\varrho^{l_\delta}_{i_1,i_2,\hdots,i_r}x_{j_{i_1}}x_{j_{i_2}}\cdots x_{j_{i_r}}x_{l_\delta}
 \end{split}
\end{equation}
From the above expression, it is clear that $\displaystyle\sum_{r=1}^k\sum_{0\leq i_1<i_2<\cdots<i_r<k}\!\!\!\!\!\!\!\!\!\!\!\!\!\varrho^{l_\delta}_{i_1,i_2,\hdots,i_r}x_{j_{i_1}}x_{j_{i_2}}\cdots x_{j_{i_r}}x_{l_\delta}$
represents the linear combination of $2^k-1$ above listed monomials with constant coefficients $\varrho^{l_\delta}_{i_1,i_2,\hdots,i_r}$ which
is the coefficient of the monomial $x_{j_{i_1}}x_{j_{i_2}}\cdots x_{j_{i_r}}x_{l_\delta}$ ($r=1,2,\hdots,k$, $0\leq i_1<i_2<\cdots<i_r<k$). It
is also noted that $\displaystyle\sum_{r=1}^k\sum_{0\leq i_1<i_2<\cdots<i_r<k}\!\!\!\!\!\!\!\!\!\!\!\!\!\varrho^{l_\delta}_{i_1,i_2,\hdots,i_r}x_{j_{i_1}}x_{j_{i_2}}\cdots x_{j_{i_r}}$
is the variable coefficient, of $x_{l_\delta}$, depends on the variables $x_{j_0},x_{j_1},\hdots,x_{j_{k-1}}$ and it is
denoted by $L^{l_\delta}_{\textbf{x}_J}$ in \textit{Theorem} 1.

Therefore,
the term $$\displaystyle\sum_{\delta=1}^p\sum_{r=1}^k\sum_{0\leq i_1<i_2<\cdots<i_r<k}\!\!\!\!\!\!\!\!\!\!\!\!\!\varrho^{l_\delta}_{i_1,i_2,\hdots,i_r}x_{j_{i_1}}x_{j_{i_2}}\cdots x_{j_{i_r}}x_{l_\delta}$$
present in $f$ produces $L^{l_\delta}_{\textbf{x}_J}$ for $\delta=1,2,\hdots,p$.

The term $\displaystyle\sum_{r=2}^k\sum_{0\leq i_1<i_2<\cdots<i_r<k}\!\!\!\!\!\!\!\!\!\!\!\!\!\alpha_{i_1,i_2,\hdots,i_r}x_{j_{i_1}}x_{j_{i_2}}\cdots x_{j_{i_r}}$
presents in $f$ represents the linear combination of the monomials of degree $2$ to $k$ over the variables $x_{j_0},x_{j_1},\hdots,x_{j_{k-1}}$
with constant coefficients. The term $\alpha_{i_1,i_2,\hdots,i_r}$ ($r=2,3,\hdots,k, 0\leq i_1<i_2<\cdots<i_r<k$) represents the
coefficient of the monomial $x_{j_{i_1}}x_{j_{i_2}}\cdots x_{j_{i_r}}$.

$\sum_{i=0}^{m-1}g_ix_i+g'$ represents the linear combination of the monomials of degree $0$ to $1$ over the variables
$x_0,x_1,\hdots,x_{m-1}$ with constant coefficients $g_i,g'$, where
$i=1,2,\hdots,m-1$.

Below, we have provided an example to illustrate the GBF given in (\ref{cgbf}).
\begin{example}
Let $f$ be a GBF of $6$ variables over $\mathbb{Z}_4$ given by
\begin{equation}
\begin{split}
 f=&2\left(x_0x_1(x_2x_4+x_4x_3+x_3x_5)\right.\\&\left.+(1-x_0)(1-x_1)(x_2x_3+x_3x_4+x_4x_5)\right.\\&\left.+x_0(1-x_1)(x_2x_4\!+\!x_4x_5)
 \!\!+\!\!(1-x_0)x_1(x_2x_3\!+\!x_3x_5)\right)\\&+(x_0+x_0x_1)x_3+3x_0x_1+2x_1x_4+x_0+2x_3+2.
 \end{split}
\end{equation}
The above given function can be obtained from (\ref{cgbf}) by setting
$k=2,p=2,j_0=0,j_1=1,l_1=3,l_2=4,S_M=\{(0,0),(1,1)\},S_{N_1}=\{(1,0)\},S_{N_2}=\{(0,1)\},(\pi_{(0,0)}(0),\pi_{(0,0)}(1),\pi_{(0,0)}(2),
\pi_{(0,0)}(3))=(2,3,4,5),(\pi_{(1,1)}(0),\pi_{(1,1)}(1),\pi_{(1,1)}(2),
\pi_{(1,1)}(3))=(2,4,3,5),(\pi^1_{(1,0)}(0),\pi^1_{(1,0)}(1),\pi^1_{(1,0)}(2))=(2,4,5),\\(\pi^2_{(0,1)}(0),\pi^2_{(0,1)}(1),\pi^2_{(0,1)}(2))=(2,3,5)$,
$\varrho_0^3=1,\varrho_1^3=0,\varrho_{0,1}^3=1,\varrho_0^4=0,\varrho_1^4=2,\varrho_{0,1}^4=0,\alpha_{0,1}=3,g_0=1,g_3=2,,g'=2,g_1=g_2=g_4=g_5=0$.
We can easily varify that $G(f\arrowvert_{(x_0,x_1)=(0,0)})$ and $G(f\arrowvert_{(x_0,x_1)=(1,1)})$ are paths over the vertices
$x_2,x_3,x_4,x_5$ and the paths are identified by the quadratic forms $x_2x_3+x_3x_4+x_4x_5$ and $x_2x_4+x_4x_3+x_4x_5$, respectively.
We can also varify that $G(f\arrowvert_{(x_0,x_1)=(1,0)})$ contains a path which is identified by the quadratic form $x_2x_4+x_4x_5$ and
one isolated vertex $x_3$. Similarly, $G(f\arrowvert_{(x_0,x_1)=(0,1)})$ contains a path which is identified by the quadratic form
$x_2x_3+x_3x_5$ and one isolated vertex $x_4$.

From the expression of the GBF $f$, it is also clear that the only term associated with $x_0$, $x_1$ and $x_3$ is given by $x_0+x_0x_1$.
Hence, $L_{\textbf{x}_J}^{l_1}=L_{(x_0,x_1)}^3=x_0+x_0x_1$. Similarly, $L_{\textbf{x}_J}^{l_2}=L_{(x_0,x_1)}^4=2x_1$.

{From (\ref{deter_ev}), we have
\begin{equation}
\begin{split}
 e_2&=x_{\pi_{(0,0)}(0)}(1-x_0)(1-x_1)+x_{\pi_{(1,1)}(0)}x_0x_1\\&+x_{\pi^1_{(1,0)}(0)}x_0(1-x_1)+x_{\pi^2_{(0,1)}(0)}(1-x_0)x_1\\
    &=x_2(1-x_0)(1-x_1)+x_2x_0x_1+x_2x_0(1-x_1)\\&+x_2x_1(1-x_0).
    \end{split}
\end{equation}}

\end{example}

We illustrate \textit{Theorem} \ref{Theorem1} by the example given below.
\begin{example}\label{examr2_2}
Let $f$ be a GBF of $5$ variables over $\mathbb{Z}_4$ given by
 \begin{equation}\label{example11_them11}
 \begin{split}
  f&=2x_1(x_0x_2+x_2x_4+x_4x_3)+2(1-x_1)(x_2x_0+x_0x_4)\\&+3x_1x_3+x_0+2x_1+1.
  \end{split}
 \end{equation}
The above given function can be obtained from (\ref{cgbf}) by setting
$k=1,p=1,j_0=1,l_1=3,S_M=\{1\},S_{N_1}=\{0\},
\left(\pi_{(1)}(0),\pi_{(1)}(1),\pi_{(1)}(2),\pi_{(1)}(3)\right)=\left(0,2,4,3\right),\left(\pi^1_{(0)}(0),
\pi^1_{(0)}(1),\pi^1_{(0)}(2)\right)=\left(2,0,4\right),\varrho_0^3=3,g_0=1,g_1=2,g_2=g_3=g_4=0$, and $g'=1$.
From (\ref{example11_them11}), we have
\begin{equation}
 \begin{split}
  f\arrowvert_{x_1=0}&=2(x_2x_0+x_0x_4)+x_0+1,\\
  f\arrowvert_{x_1=1}&=2(x_0x_2+x_2x_4+x_4x_3)+3x_3+x_0+3.
 \end{split}
\end{equation}
Hence, $G( f\arrowvert_{x_1=1})$ is a path over the vertices $x_0,x_2,x_3,x_4$ and $G( f\arrowvert_{x_1=0})$ contains a path
over the vertices $x_0,x_2,x_4$ and one isolated vertex $x_3$. Fig. 1 (a) and Fig. 1 (b) represent $G(f\big\arrowvert_{x_1=1})$
and $G(f\big\arrowvert_{x_1=0})$, respectively.
Using (\ref{deter_ev}), we have
\begin{equation}
 e_2=x_0x_1+x_2(1-x_1).
\end{equation}
Therefore, $e_2\arrowvert_{x_1=0}=x_2$ which is a end vertex of the path in $G(f\arrowvert_{x_1=0})$ and $e_2\arrowvert_{x_1=1}$ gives
the end vertex $x_0$ of the path $G(f\arrowvert_{x_1=1})$. Following \textit{Theorem} 1, we obtain the set $S$ corresponding to the GBF $f$
as follows:
\begin{equation}
\begin{split}
S&=\left \{f+2\left(d_0x_1+d''e_2\right): d_0\in \{0,1\}, d''\in\{0,1\}\right\}\\
&=\begin{bmatrix}
   12301032122310211030121010011221\\
   12121010120110031012123210231203\\
   12323230122132231032301210033023\\
   12103212120332011010303010213001\\
  \end{bmatrix}
  \end{split}
\end{equation}
\begin{figure}[!t]
\centering
\includegraphics[height=4cm]{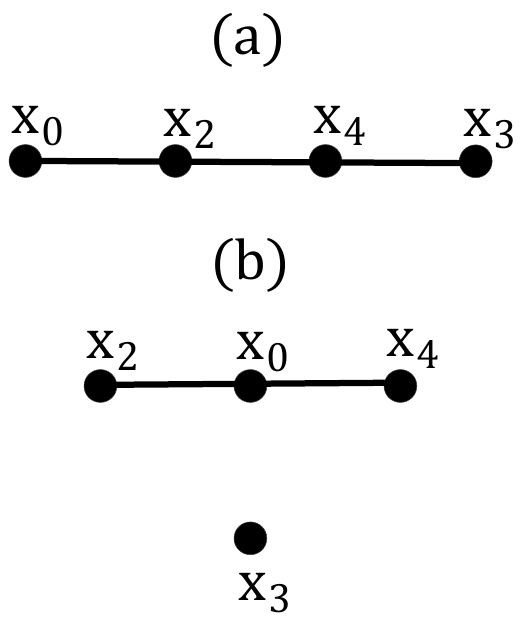}
\caption{The $G(f\big\arrowvert_{x_1=1})$ and $G(f\big\arrowvert_{x_1=0})$ of \textit{Example \ref{examr2_2}}.}
\end{figure}
In the expression of the GBF $f$, the only term associated with the restricting variable $x_1$ and $x_{l_1}$ ($=x_3$) is $3x_1x_3$. Therfore,
following \textit{Theorem} 1, we have $L_{\textbf{x}_J}^{l_1}=L^{3}_{x_1}=3x_1$ and the AACF of $S$ is given by
\begin{equation}
 \begin{split}
  A(S)(\tau)=\begin{cases}
              128, &\tau=0,\\
              32\omega_4^{L_0^3},&\tau=8,\\
              32\omega_4^{-L_0^3},&\tau=-8,\\
              0,&\textnormal{otherwise}.
             \end{cases}
 \end{split}
\end{equation}
Since, $L_0^3=0$, we have
\begin{equation}
 \begin{split}
  A(S)(\tau)=\begin{cases}
              128, &\tau=0,\\
              32,&\tau=\pm 8,\\
              0,&\textnormal{otherwise}.
             \end{cases}
 \end{split}
\end{equation}
\end{example}

\begin{remark}\label{remark1}
Let $f$ be a quadratic GBF with the property that for all $\textbf{c}\in \{0,1\}^k$, $G(f\big\arrowvert_{\textbf{x}_J=\textbf{c}})$ is a path in $m-k$
vertices. Then from \textit{Therorem} \ref{Theorem1}, we have $M=2^{k}$ and
\begin{equation}
 \begin{split}
  A(S)(\tau)=\begin{cases}
        2^{m+k+1}, & \tau=0,\\
        0, & \textnormal{otherwise.}
       \end{cases}
 \end{split}
\end{equation}
Hence, $S$ is a CS of size $2^{k+1}$ and therefore, Paterson's construction \cite[Th. 12]{pater2000} turns to be a special case of our proposed one.
\end{remark}
\begin{remark}
From \textit{Remark} \ref{remark1},
for $k=0$, $S$ is a CS of size $2$, i.e., $S$ is a GCP and thus the GDJ code in \cite{Davis1999} is also a special case of \textit{Theorem} \ref{Theorem1}.
\end{remark}
\begin{remark}
Let $f$ be a quadratic GBF with the property that for all $\textbf{c}\in \{0,1\}^k$, $G(f\big\arrowvert_{\textbf{x}_J=\textbf{c}})$ contains a path in $m-k-1$
vertices and one isolated vertex $x_{l_1}$. We also assume that all edges in the original graph between the isolated vertex and the $k$ deleted vertices
are weighted by $q/2$. Then, from \textit{Therorem} \ref{Theorem1}, we have $N_1=2^{k}$, $S_{N_1}=\{0,1\}^k$,
$L^{l_1}_{\textbf{c}}=\frac{q}{2}\sum_{\alpha=0}^{k-1}c_\alpha$, and
\begin{equation}
 \begin{split}
  A(S)(\tau)&=\begin{cases}
        2^{m+k+1}, & \tau=0,\\
        \omega_q^{g_{l_1}}2^{m+k}\displaystyle\sum_{\textbf{c}\in S_{N_1}}\omega_q^{L^{l_1}_{\textbf{c}}}, & \tau=2^{l_1},\\
        \omega_q^{-g_{l_1}}2^{m+k}\displaystyle\sum_{\textbf{c}\in S_{N_1}}\omega_q^{-L^{l_1}_{\textbf{c}}}, & \tau=-2^{l_1},\\
        0, & \textnormal{otherwise},
       \end{cases}\\
       &=\begin{cases}
        2^{m+k+1}, & \tau=0,\\
        0, & \textnormal{otherwise.},
       \end{cases}
 \end{split}
\end{equation}
Therefore, $\psi(f)$ lies in a CS of size $2^{k+1}$ and the result given by Paterson in \cite[Th. 24]{pater2000} turns to be a special case of
\textit{Theorem} \ref{Theorem1}.
\end{remark}
\begin{remark}
Let $f$ be a GBF with the property that for all $\textbf{c}\in \{0,1\}^k$, $G(f\big\arrowvert_{\textbf{x}_J=\textbf{c}})$ is a path in $m-k$
vertices. Then from \textit{Therorem} \ref{Theorem1}, we have $M=2^{k}$ and
\begin{equation}\label{stinspecial}
 \begin{split}
  A(S)(\tau)=\begin{cases}
        2^{m+k+1}, & \tau=0,\\
        0, & \textnormal{otherwise.}
       \end{cases}
 \end{split}
\end{equation}
From (\ref{stinspecial}), it is clear that $\psi(f)$ lies in a CS of size $2^{k+1}$ and hence the PMEPR of $\psi(f)$ is atmost
$2^{k+1}$. Therefore, the result given by Schmidt in \cite[Th. 5]{Schmid2007} is a special case of \textit{Theorem} \ref{Theorem1}.
\end{remark}
\section{Proposed Constructions of Complementary Sequences with Low PMEPR}
In this section, we present two constructions of CSs which are derived from \textit{Theorem} \ref{Theorem1} to provide tighter PMEPR upper bound than the PMEPR bound introduced in Schmidt's construction
\cite[Th. 5]{Schmid2007}.
\begin{corollary}\label{corr_1}
Let $f$ be a GBF as defined in \textit{Theorem} \ref{Theorem1} with the property that $N_i\equiv 0(\!\!\!\!\mod 2)$ ($i=1,2,\hdots,p$) and there exist
$N_i/2$ number of $\textbf{c}$ in $S_{N_i}$ for which $L^{l_i}_{\textbf{c}}\equiv 0 (\!\!\!\!\mod q)$, and
$L^{l_i}_{\textbf{c}}\equiv \frac{q}{2} (\!\!\!\!\mod q)$
for the rest $N_i/2$ number of $\textbf{c}$ in $S_{N_i}$.
Then for any choice of $g_i$, $g'\in$ $\mathbb{Z}_q$,
\begin{equation}
 \left \{f+\frac{q}{2}\left(\textbf{d}\cdot\textbf{x}_J+d''e_2\right):\textbf{d}\in \{0,1\}^k,d''\in\{0,1\}\right\},
\end{equation}
is a CS of size $2^{k+1}$.
\begin{IEEEproof}
Let
  \begin{equation}
      \begin{split}
        S=\left \{f+\frac{q}{2}\left(\textbf{d}\cdot\textbf{x}_J+d''e_2\right):\textbf{d}\in \{0,1\}^k,d''\in\{0,1\}\right\}.
        \end{split}
     \end{equation}
By \textit{Theorem 1}, we have 
\begin{equation}\label{autocorrelation_comple_1}
 \begin{split}
   A(S)(\tau)\!\!=\!\!\begin{cases}
        2^{m+k+1},& \tau=0,\\
        \omega_q^{g_{l_i}}2^m\displaystyle\sum_{\textbf{c}\in S_{N_i}}\omega_q^{L^{l_i}_{\textbf{c}}}, & \tau\!=\!2^{l_i},i\!\!=\!\!1,2,\hdots,p,\\
        \omega_q^{-g_{l_i}}2^m\displaystyle\sum_{\textbf{c}\in S_{N_i}}\omega_q^{-L^{l_i}_{\textbf{c}}}, & \tau=-2^{l_i},i\!\!=\!\!1,2,\!\hdots\!,p,\\
        0, & \textnormal{otherwise}.
       \end{cases}
 \end{split}
\end{equation}
Since there exist $N_i/2$ number of $\textbf{c}$ in $S_{N_i}$ for which $L^{l_i}_{\textbf{c}}\!\!\equiv\!\! 0 (\!\!\!\!\mod q)$, $\omega_q^{L^{l_i}_{\textbf{c}}}$
takes the value $1$ for $N_i/2$ times. Similarly, $\omega_q^{L^{l_i}_{\textbf{c}}}$ takes the value $-1$ for $N_i/2$ times.
Therefore, $\sum_{\textbf{c}\in S_{N_i}}\omega_q^{L^{l_i}_{\textbf{c}}}=0$. In the same way, we can show
that $\sum_{\textbf{c}\in S_{N_i}}\omega_q^{-L^{l_i}_{\textbf{c}}}=0$. Hence, from (\ref{autocorrelation_comple_1}), we have
\begin{equation}\label{comple_11}
 \begin{split}
  A(S)(\tau)=\begin{cases}
        2^{m+k+1}, & \tau=0,\\
        0, & \textnormal{otherwise}.
       \end{cases}
 \end{split}
\end{equation}
From (\ref{comple_11}), we have $S$ is a CS of size $2^{k+1}$ and hence at most PMEPR of each  sequences
lying in $S$ is $2^{k+1}$ \cite{pater2000}.
\end{IEEEproof}
\end{corollary}
\begin{remark}[Explicit Form of GBFs as Defined in \textit{Corollary} \ref{corr_1}]
To construct the GBFs as defined in \textit{Corollary} \ref{corr_1}, we only need to take care of the following term in (\ref{cgbf}):
$$\displaystyle\sum_{\delta=1}^p\sum_{r=1}^k\sum_{0\leq i_1<i_2<\cdots<i_r<k}\!\!\!\!\!\!\!\!\!\!\!\!\!\varrho^{l_\delta}_{i_1,i_2,\hdots,i_r}
x_{j_{i_1}}x_{j_{i_2}}\cdots x_{x_{i_r}}x_{l_\delta},$$ or $\sum_{\delta=1}^pL^{l_\delta}_{\mathbf{x}_J}x_{l_\delta}$. In this \textit{Remark}, we define
$L^{l_\delta}_{\mathbf{x}_J}$, so that the GBFs associated with $L^{l_\delta}_{\mathbf{x}_J}$, meet the condition given in \textit{Corollary} \ref{corr_1}.
To define $L^{l_\delta}_{\mathbf{x}_J}$, first we need to define
some vectors which are as follows: $\textbf{c}^{l_\delta}_{\phi_t}=(c^{l_\delta}_{0,\phi_t},c^{l_\delta}_{1,\phi_t},\hdots,c^{l_\delta}_{k-1,\phi_t})\in S_{N_\delta}$,
where $t=1,2,\hdots, N_\delta/2$, $\delta=1,2,\hdots,p$. Therefore, $\textbf{c}^{l_\delta}_{\phi_1}, \textbf{c}^{l_\delta}_{\phi_2},\hdots,\textbf{c}^{l_\delta}_{\phi_{N_\delta/2}}$
are any $N_\delta/2$ distinct elements in $S_{N_\delta}$. Let us define
\begin{equation}\label{lterm}
L^{l_\delta}_{\mathbf{x}_J}=\frac{q}{2}\displaystyle\sum_{t=1}^{N_\delta/2}\prod_{\alpha=0}^{k-1}x_{j_\alpha}^{c^{l_\delta}_{\alpha,\phi_t}}(1-x_{j_\alpha})^{(1-c^{l_\delta}_{\alpha,\phi_t})}.
\end{equation}
From the above equation, it is clear that $L^{l_\delta}_{\mathbf{x}_J}=1$ for $\mathbf{x}_J=\textbf{c}^{l_\delta}_{\phi_t}$, $t=1,2,\hdots,N_\delta/2$ and
for the remaining of $N_\delta/2$ elements in $S_{N_\delta}$, $L^{l_\delta}_{\mathbf{x}_J}=0$. Therefore, the GBFs whose $L^{l_\delta}_{\mathbf{x}_J}$ terms
are as defined as in (\ref{lterm}), satisfy the conditions given in \textit{Corollary} \ref{corr_1}.
\end{remark}
\begin{remark}
The construction of CSs given in \textit{Corollary \ref{corr_1}} is based on GBFs of any order.
It is observed that
\textit{Corollary \ref{corr_1}} can provide tighter upper bound of PMEPR than that given by Schmidt \cite[Th. 5]{Schmid2007} for a sequence
corresponding to a GBF which satisfies the property given in \textit{Corollary} \ref{corr_1}. Below,
we present an example to illustrate \textit{Corollary \ref{corr_1}}.
\end{remark}

\begin{example}
 Let $f$ be a GBF of $5$ variables over $\mathbb{Z}_4$, given by
 \begin{equation}
 \begin{split}
  f&=2\left(x_0x_1x_2+x_0x_1x_3+x_1x_3+x_3x_2+x_0x_4\right)\\&+x_1+2x_2+2x_3+2x_4+3\\
  &\equiv 2x_0(x_3x_2+x_2x_1)+2(1-x_0)(x_2x_3+x_3x_1)\\&+2x_0x_4+x_1+2x_2+2x_3+2x_4+3.
  \end{split}
 \end{equation}
The GBF $f$ can be obtained from (\ref{cgbf}) by substituting $k=1$, $p=1$, $M=0$, $N_1=2$, $S_{N_1}=\{0,1\}$, $j_0=0$,
$\left(\pi_{(0)}^1(0),\pi_{(0)}^1(1),\pi_{(0)}^1(2)\right)=\left(2,3,1\right)$,
$\left(\pi_{(1)}^1(0),\pi_{(1)}^1(1),\pi_{(1)}^1(2)\right)=\left(3,2,1\right)$, $l_1=4$,
$\varrho_0^4=2$, $g_0=0$, $g_1=1$, $g_2=g_3=g_4=2$, and $g'=3$.

From the GBF $f$, we obtain the restricted Boolean functions as follows.
\begin{equation}\label{exam_corr_1}
\begin{split}
 f\big\arrowvert_{x_0=0}&=2(x_2x_3+x_3x_1)+x_1+2x_2+2x_3+2x_4+3,\\
 f\big\arrowvert_{x_0=1}&=2(x_3x_2+x_2x_1)+x_1+2x_2+2x_3+3.
 \end{split}
\end{equation}
From (\ref{exam_corr_1}), it is observed that $G(f\big\arrowvert_{x_0=0})$ and $G(f\big\arrowvert_{x_0=1})$ both contain a path over
the vertices $x_1,x_2,x_3$ and one isoltaed vertex $x_4$.

We can easily varify that $2x_0x_4$ is the only term present in $f$ and associated with the restricting variable $x_0$ and isolated vertex $x_4$.
Therefore, $L_{x_0}^4=2x_0$, $L^4_0=0$, and $L^4_1=2$. From (\ref{deter_ev}), we have $e_2=x_2(1-x_0)+x_3x_0$.
Using \textit{Corollary \ref{corr_1}},
\begin{equation}
 \begin{split}
  S\!=\!&\left\{2\left(x_0x_1x_2\!+\!x_0x_1x_3\!+\!x_1x_3\!+\!x_3x_2\!+\!x_0x_4\right)\!+\!x_1\!+\!2x_2\right.\\ &\left.+2x_3+2x_4+3+2(d_0x_0+d''e_2):
  d_0,d''\in \{0,1\} \right\}\\
  =&\begin{bmatrix}
     3     3     0     0     1     1     2     0     1     1     0     2     1     1     0     0     1     3     2     0     3     1     0     0     3     1     2     2     3     1     2     0\\
     3     1     0     2     1     3     2     2     1     3     0     0     1     3     0     2     1     1     2     2     3     3     0     2     3     3     2     0     3     3     2     2\\
     3     3     0     0     3     1     0     0     1     3     0     0     3     3     2     2     1     3     2     0     1     1     2     0     3     3     2     0     1     3     0     2\\
     3     1     0     2     3     3     0     2     1     1     0     2     3     1     2     0     1     1     2     2     1     3     2     2     3     1     2     2     1     1     0     0
    \end{bmatrix}.
 \end{split}
\end{equation}
is a CS of size $4$. Therefore, the PMEPR of $\psi(f)$ is at most $4$ and from Schmidt's construction, the PMEPR upper bound of $\psi(f)$ is $8$.
\end{example}
\begin{corollary}\label{corr_2}
Let $f$ be a GBF as defined in \textit{Theorem \ref{Theorem1}} and unlike the GBF as defined in \textit{Corollary} \ref{corr_1}. Then for any
 choice of $g_i$, $g'\in$ $\mathbb{Z}_q$,
\begin{equation}
\begin{split}
 \left \{f\!+\!\frac{q}{2}\left(\textbf{d}\!\cdot\!\textbf{x}_J\!+\!d'\sum_{i=1}^px_{l_i}\!+\!d''e_2\right):\right.\\~~~~~~~~~~~\left.\textbf{d}\!\in\! \{0,1\}^k,d',d''\!\in\!\{0,1\}\right\},
\end{split}
 \end{equation}
is a CS of size $2^{k+2}$ with at most PMEPR $2^{k+2}-2M$.
\begin{IEEEproof}
 The set $S$ can be expressed as $S=S_1\cup S_2$, where
 \begin{equation}
 \begin{split}
 S_1&\!=\!\left \{f\!+\!\frac{q}{2}\left(\textbf{d}\!\cdot\!\textbf{x}_J\!+\!d''e_2\right):\textbf{d}\!\in\! \{0,1\}^k,d''\!\in\!\{0,1\}\right\},\\
 S_2&\!=\!\left \{f\!+\!\frac{q}{2}\left(\textbf{d}\!\cdot\!\textbf{x}_J\!+\!\sum_{i=1}^px_{l_i}\!+\!d''e_2\right)\!:\!\textbf{d}\!\in\! \{0,1\}^k,d''\!\in\!\{0,1\}\right\}.
 \end{split}
 \end{equation}
 By \textit{Theorem \ref{Theorem1}}, we have
 \begin{equation}\label{T2A1}
 \begin{split}
  A(S_1)(\tau)\!\!=\!\!\begin{cases}
        2^{m+1}\displaystyle\sum_{i=1}^pN_i+2^{m+1}M, \tau=0,\\
        \omega_q^{g_{l_i}}2^m\displaystyle\sum_{c\in S_{N_i}}\omega_q^{L^{l_i}_{\textbf{c}}},~~~~~~\tau\!=\!2^{l_i},i\!\!=\!\!1,2,\hdots,p,\\
        \omega_q^{-g_{l_i}}2^m\displaystyle\sum_{c\in S_{N_i}}\omega_q^{-L^{l_i}_{\textbf{c}}},~~\tau\!=\!-2^{l_i},i\!\!=\!\!1,2,\!\hdots\!,p,\\
        0,~~~~~~~~~~~~~~~~~~~~~~~~~~ \textnormal{otherwise},
       \end{cases}
 \end{split}
\end{equation}
and
\begin{equation}\label{T2A2}
 \begin{split}
  A(S_2)(\tau)\!\!&=\!\!\begin{cases}
        2^{m+1}\displaystyle\sum_{i=1}^pN_i+2^{m+1}M, \tau=0,\\
        \omega_q^{\frac{q}{2}+g_{l_i}}2^m\displaystyle\sum_{c\in S_{N_i}}\omega_q^{L^{l_i}_{\textbf{c}}},\tau\!=\!2^{l_i},i\!\!=\!\!1,2,\hdots,p,\\
        \omega_q^{-(\frac{q}{2}+g_{l_i})}2^m\!\!\!\!\displaystyle\sum_{c\in S_{N_i}}\!\!\!\!\!\omega_q^{-L^{l_i}_{\textbf{c}}}, \tau\!=\!-2^{l_i},i\!\!=\!\!1,2,\!\hdots\!,p,\\
        0, ~~~~~~~~~~~~~~~~~~~~~~~~~\textnormal{otherwise}.
       \end{cases}\\
       \!\!&=\!\!\begin{cases}
        2^{m+1}\displaystyle\sum_{i=1}^pN_i+2^{m+1}M, \tau=0,\\
        -\omega_q^{g_{l_i}}2^m\displaystyle\sum_{c\in S_{N_i}}\omega_q^{L^{l_i}_{\textbf{c}}},~~~~ \tau\!=\!2^{l_i},i\!\!=\!\!1,2,\hdots,p,\\
        -\omega_q^{-g_{l_i}}2^m\displaystyle\sum_{c\in S_{N_i}}\omega_q^{-L^{l_i}_{\textbf{c}}},  \tau\!=\!-2^{l_i},i\!\!=\!\!1,2,\!\hdots\!,p,\\
        0, ~~~~~~~~~~~~~~~~~~~~~~~~~~\textnormal{otherwise}.
       \end{cases}
 \end{split}
\end{equation}
From (\ref{T2A1}) and (\ref{T2A2}), we have
\begin{equation}
\begin{split}
 A(S_1)(\tau)+A(S_2)(\tau)=\begin{cases}
        2^{m+k+2}, & \tau=0,\\
        0, & \textnormal{otherwise}.
       \end{cases}
       \end{split}
\end{equation}
Therefore, $S$ is a CS of size $2^{k+2}$. Let us assume that $S_1=\{\textbf{a}^0,\textbf{a}^1,\hdots,\textbf{a}^{2^{k+1}-1}\}$.
From (\ref{pmeprdef}) and
(\ref{T2A1}), we have
\begin{equation}\label{pmepr111}
 \begin{split}
  P(\textbf{a}^\alpha)(t)&\leq \displaystyle\sum_{\beta=0}^{2^{k+1}-1}P(\textbf{a}^\beta)(t)\\
                    &\leq 2^{m+k+1}+2^{m}\sum_{i=1}^p\sum_{c\in S_{N_i}}\left[|\omega_q^{L^{l_i}_{\textbf{c}}}|+|\omega_q^{-L^{l_i}_{\textbf{c}}}|\right]\\
                    &=2^{m+k+1}+2^{m+1}\sum_{i=1}^p N_i,
 \end{split}
\end{equation}
where $\alpha=0,1,\hdots,2^{k+1}-1$.
From (\ref{pmepr111}), we have
\begin{equation}\label{pmmmer}
\begin{split}
\frac{P(\textbf{a}^\alpha)(t)}{2^m}&\leq 2^{k+1}+ 2\sum_{i=1}^p N_i \\
                                   &=2^{k+2}-2M.
\end{split}
\end{equation}
From (\ref{defipmeprnew}) and (\ref{pmmmer}), it is clear that the PMEPR of $\textbf{a}^\alpha$ is upper bounded by $2^{k+2}-2M$ for all $\alpha=0,1,\hdots,2^{k+1}-1$.
Similarly, we can show that the PMEPRs of the sequences in $S_2$ are upper bounded by $2^{k+2}-2M$.
Since $S$ is the union of sets $S_1$ and $S_2$, the PMEPR of $S$
is at most $2^{k+2}-2M$.
\end{IEEEproof}
\end{corollary}
\begin{remark}
 It is observed that \textit{Corollary} \ref{corr_2} can provide more tight upper bound of PMEPR than that of \cite[Th. 5]{Schmid2007}
 for a sequence corresponding to a GBF which satisfies the properties introduced
 in \textit{Corollary} \ref{corr_2}.
\end{remark}
\begin{example}
Let $f$ be a GBF of $5$ variables $x_0,x_1,x_2,x_3,x_4$ over $\mathbb{Z}_4$, given by
\begin{equation}
\begin{split}
 f&=2(x_0x_1x_3+x_0x_3x_4+x_1x_3+x_3x_2)\\
 &\equiv 2x_0(x_4x_3+x_3x_2)+2(1-x_0)(x_1x_3+x_3x_2).
 \end{split}
\end{equation}
The above GBF can be obtained from (\ref{cgbf}) by substituting $k=1$, $j_0=0$, $p=2$, $M=0$, $N_1=1$, $N_2=1$, $S_{N_1}=\{0\}$, $S_{N_2}=\{1\}$,
  $\left(\pi_{(0)}^1(0),\pi_{(0)}^1(1),\pi_{(0)}^1(2)\right)=(1,3,2)$,
$\left(\pi_{(1)}^2(0),\pi_{(1)}^2(1),\pi_{(1)}^2(2)\right)=(4,3,2)$, $l_1=4$, $l_2=1$,
$\varrho_0^4=0$, $\varrho_0^1=0$ and $g_0=g_1=g_2=g_3=g_4=g'=0$.

The restricted Boolean functions $f\big\arrowvert_{x_0=0}$ and $f\big\arrowvert_{x_0=1}$ are
\begin{equation}\label{exa_2_eq}
\begin{split}
 f\big\arrowvert_{x_0=0}&=2(x_1x_3+x_3x_2),\\
 f\big\arrowvert_{x_0=1}&=2(x_4x_3+x_3x_2),
 \end{split}
\end{equation}
respectively. From (\ref{exa_2_eq}), it is clear that $G(f\big\arrowvert_{x_0=0})$ contains one path over the vertices $x_1,x_2,x_3$ and $x_4$
as isolated vertex, and $G(f\big\arrowvert_{x_0=1})$ contains one path over the vertices $x_2,x_3,x_4$ and $x_1$ as isolated vertex.
We can easily verify that there is no term, present in $f$, associated with $x_0$ and isolated vertices $x_1$, $x_4$. Therefore,
$L_{x_0}^1=0$ and $L_{x_0}^4=0$. From (\ref{deter_ev}), we have $e_2=x_1(1-x_{0})+x_4x_{0}$.
Using \textit{Corollary \ref{corr_2}}, the set
\begin{equation}
\begin{split}
 &\left\{f+2\left(d_0x_0+d'(x_1+x_4)+d''e_2\right):d_0,d',d''\in\{0,1\}\right\}\\
 &=\begin{bmatrix}
     0     0     0     0     0     0     0     0     0     0     2     0     2     2     0     2     0     0     0     0     0     0     0     0     0     2    2     2     2     0     0     0\\
     0     2     0     2     0     2     0     2     0     2     2     2     2     0     0     0     0     2     0     2     0     2     0     2     0     0    2     0     2     2     0     2\\
     0     0     2     2     0     0     2     2     0     0     0     2     2     2     2     0     2     2     0     0     2     2     0     0     2     0    2     2     0     2     0     0 \\
     0     2     2     0     0     2     2     0     0     2     0     0     2     0     2     2     2     0     0     2     2     0     0     2     2     2    2     0     0     0     0     2 \\
     0     0     2     0     0     0     2     0     0     0     0     0     2     2     2     2     0     2     2     2     0     2     2     2     0     0    0     0     2     2     2     2 \\
     0     2     2     2     0     2     2     2     0     2     0     2     2     0     2     0     0     0     2     0     0     0     2     0     0     2    0     2     2     0     2     0 \\
     0     0     0     2     0     0     0     2     0     0     2     2     2     2     0     0     2     0     2     2     2     0     2     2     2     2    0     0     0     0     2     2 \\
     0     2     0     0     0     2     0     0     0     2     2     0     2     0     0     2     2     2     2     0     2     2     2     0     2     0    0     2     0     2     2     0
    \end{bmatrix}
 \end{split}
\end{equation}
is a CS of size $8$. Hence, by using \textit{Corollary \ref{corr_2}}, the PMEPR upper bound for $\psi(f)$ is
$8$ whereas Schmidt's construction provides a PMEPR upper bound of $16$.
\end{example}
\begin{example}\label{example_corr_th2}
Let $f$ be a GBF of $6$ variables over $\mathbb{Z}_4$, given by
\begin{equation}\label{example_2_eq}
\begin{split}
f=&2(x_0x_2x_3+x_0x_3x_4+x_0x_4x_5+x_0x_2x_4+x_0x_1x_4\\&+x_0x_1x_3+x_0x_3x_5+x_2x_4+x_4x_1+x_1x_3+x_3x_5)\\
\equiv & 2x_0(x_2x_3+x_3x_4+x_4x_5)\\&+2(1-x_0)(x_2x_4+x_4x_1+x_1x_3+x_3x_5).
\end{split}
\end{equation}
The above GBF can be obtained from (\ref{deter_ev}) by substituting $k=1$, $j_0=x_0$, $p=1$, $M=1$, $N_1=1$, $S_M=\{0\}$, $S_{N_1}=\{1\}$,
$\left(\pi_{(0)}(0),\pi_{(0)}(1),\pi_{(0)}(2),\pi_{(0)}(3),\pi_{(0)}(4)\right)=(2,4,1,3,5)$,
$\left(\pi_{(0)}(0),\pi_{(0)}(1),\pi_{(0)}(2),\pi_{(0)}(3)\right)=(2,3,4,5)$, $l_1=1$, $\varrho_{0}^1=0$ and $g_0=g_1=\cdots=g_5=g'=0$.

The restricted Boolean functions $f\big\arrowvert_{x_0=0}$ and $f\big\arrowvert_{x_0=1}$ are given by
\begin{equation}
\begin{split}
 f\big\arrowvert_{x_0=0}&=2(x_2x_4+x_4x_1+x_1x_3+x_3x_5),\\
 f\big\arrowvert_{x_0=1}&=2(x_2x_3+x_3x_4+x_4x_5),
 \end{split}
\end{equation}
respectively.
\begin{figure}[!t]
\centering
\includegraphics[height=7cm]{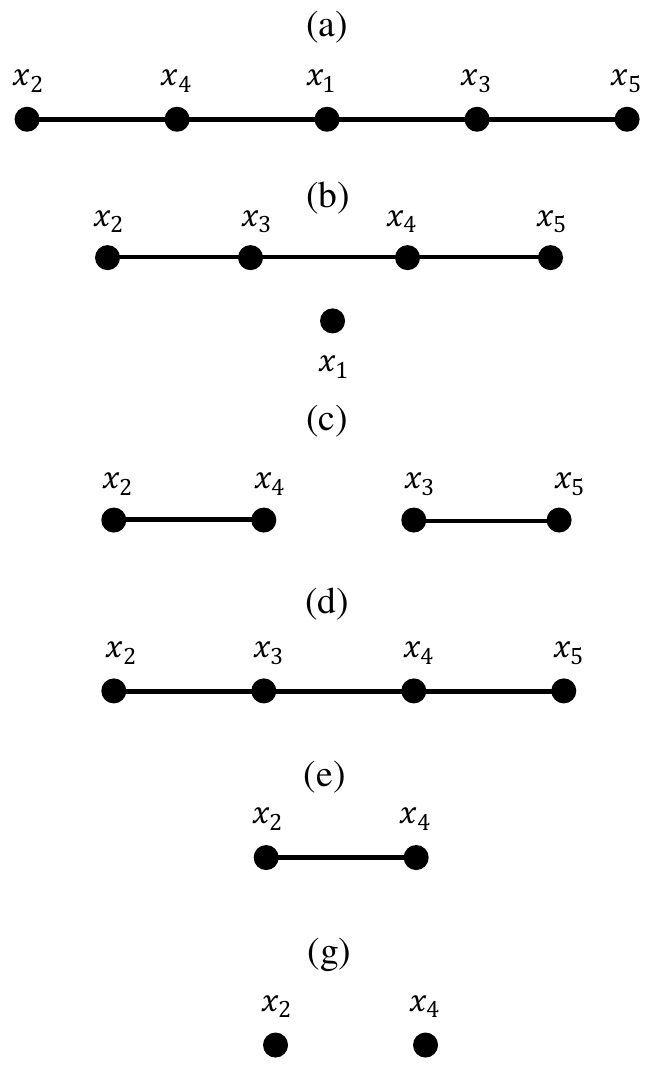}
\caption{The graphs of the restricted Boolean functions obtained from $f$.}
\end{figure}
It is clear that $G(f\big\arrowvert_{x_0=0})$ is a path and $G(f\big\arrowvert_{x_0=1})$ contains a path and the isolated vertex $x_1$.
From the expression of the GBF $f$, we can easily varify that there is no term associated with the variables $x_0$ and $x_1$. Therefore,
$L_{x_0}^1=0$. From (\ref{deter_ev}), we have $e_2=x_2(1-x_0)+x_2x_0=x_2$.

Using \textit{Corollary \ref{corr_2}}, the set
\begin{equation}
 \begin{split}
  S&=\left\{f+2(d_0x_0+d'x_1+d''x_2):
  d_0,d',d''\in \{0,1\} \right\}\\
  &\tiny=\begin{bmatrix}
0     0     0     0     0     0     0     0     0     0     2     0     0     2     2     2     0     0     2     0     2     0     0     0     0     2    0     2     2     0     2     0     0     0     0     0     0     0     0     0     2     0     0     0     2     2     0     2     0     2     2     2    2     2     0     2     2     0     2     0     0     2     0     2\\
0     2     0     2     0     2     0     2     0     2     2     2     0     0     2     0     0     2     2     2     2     2     0     2     0     0    0     0     2     2     2     2     0     2     0     2     0     2     0     2     2     2     0     2     2     0     0     0     0     0     2     0    2     0     0     0     2     2     2     2     0     0     0     0\\
0     0     2     2     0     0     2     2     0     0     0     2     0     2     0     0     0     0     0     2     2     0     2     2     0     2    2     0     2     0     0     2     0     0     2     2     0     0     2     2     2     0     2     2     2     2     2     0     0     2     0     0    2     2     2     0     2     0     0     2     0     2     2     0\\
0     2     2     0     0     2     2     0     0     2     0     0     0     0     0     2     0     2     0     0     2     2     2     0     0     0    2     2     2     2     0     0     0     2     2     0     0     2     2     0     2     2     2     0     2     0     2     2     0     0     0     2    2     0     2     2     2     2     0     0     0     0     2     2\\
0     0     0     0     2     2     2     2     0     0     2     0     2     0     0     0     0     0     2     0     0     2     2     2     0     2    0     2     0     2     0     2     0     0     0     0     2     2     2     2     2     0     0     0     0     0     2     0     0     2     2     2    0     0     2     0     2     0     2     0     2     0     2     0\\
0     2     0     2     2     0     2     0     0     2     2     2     2     2     0     2     0     2     2     2     0     0     2     0     0     0    0     0     0     0     0     0     0     2     0     2     2     0     2     0     2     2     0     2     0     2     2     2     0     0     2     0    0     2     2     2     2     2     2     2     2     2     2     2\\
0     0     2     2     2     2     0     0     0     0     0     2     2     0     2     2     0     0     0     2     0     2     0     0     0     2    2     0     0     2     2     0     0     0     2     2     2     2     0     0     2     0     2     2     0     0     0     2     0     2     0     0    0     0     0     2     2     0     0     2     2     0     0     2\\
0     2     2     0     2     0     0     2     0     2     0     0     2     2     2     0     0     2     0     0     0     0     0     2     0     0    2     2     0     0     2     2     0     2     2     0     2     0     0     2     2     2     2     0     0     2     0     0     0     0     0     2    0     2     0     0     2     2     0     0     2     2     0     0
\end{bmatrix}.
\end{split}
\end{equation}
is a complementary set of size $8$ and the PMEPR upper bound of the sequences lying in $S$ is $6$.
The $G(f\big\arrowvert_{x_0=0})$ and $G(f\big\arrowvert_{x_0=1})$ are represented by Fig. 2 (a) and Fig. 2 (b) respectively. Since, $G(f\big\arrowvert_{x_0=1})$
contains the isolated vertex $x_1$, Schmidt's construction suggests to delete the isolated vertex $x_1$. After deleting the isolated vertex or restricting
$x_1$, we obtain the following restricted Boolean functions $f\big\arrowvert_{(x_0,x_1)=(0,0)}$, $f\big\arrowvert_{(x_0,x_1)=(0,1)}$,
$f\big\arrowvert_{(x_0,x_1)=(1,0)}$ and $f\big\arrowvert_{(x_0,x_1)=(1,1)}$. The $G(f\big\arrowvert_{(x_0,x_1)=(0,0)})$, $G(f\big\arrowvert_{(x_0,x_1)=(0,1)})$,
are represented by Fig. 2 (c) and $G(f\big\arrowvert_{(x_0,x_1)=(1,0)})$, $G(f\big\arrowvert_{(x_0,x_1)=(1,1)})$ are represented
by Fig. 2 (d). Again, deletion needs to be performed by following Scmidt's construction. After performing another deletion of vertices, the resulting
graphs of restricted Boolean functions will be represented by Fig. 2 (e) and Fig. 2 (g). The deletion process can continue until the graph of every
restricted Boolean function is a path or consists of a single vertex.

Therefore, from Schmidt's construction, the PMEPR upper bound of $\psi(f)$ is $64$ whereas from \textit{Corollary \ref{corr_2}},
the PMEPR upper bound of $\psi(f)$
is $6$. Note that $4$-PSK is considered in this example.
\end{example}
\section{Graphical Analysis of the Proposed Constructions}
In this section, we interpret our proposed construction with graphical analysis.

A graph can be represented by a pair of sets $(V,E)$, where $V$ is the set of verices and $E$ is the set of edges present in the graph.
As an example, the graph given in Fig. 1 (a) can also be expressed by $(V,E)$, where $V=\{x_1,x_2,x_3\}$ and $E=\{x_1x_3, x_2x_3\}$. The term
$x_1x_3$ represents the edge between the vertices $x_1$ and $x_3$. Similarly, $x_2x_3$ represents the edge between the vertices $x_2$ and $x_3$. We say
a graph $(V,E)$ is a path if the edges in $E$ form a path over all the vertices presented in $V$. If there exist some vertices in $V$ which are
not associated with any edges presented in $E$, we call them isolated vertices in the graph $(V,E)$. As an example, in Fig. 1 (b), $V=\{x_1,x_2,x_3\}$ and
$E=\{x_1x_2\}$, where the set $E$ does not contain any such edges which include the vertex $x_3$. Hence, for Fig. 1 (b), we call $(V,E)$, a graph containing
a path and an isolated vertex.
\begin{figure}[!h]
\centering
\includegraphics[height=6cm]{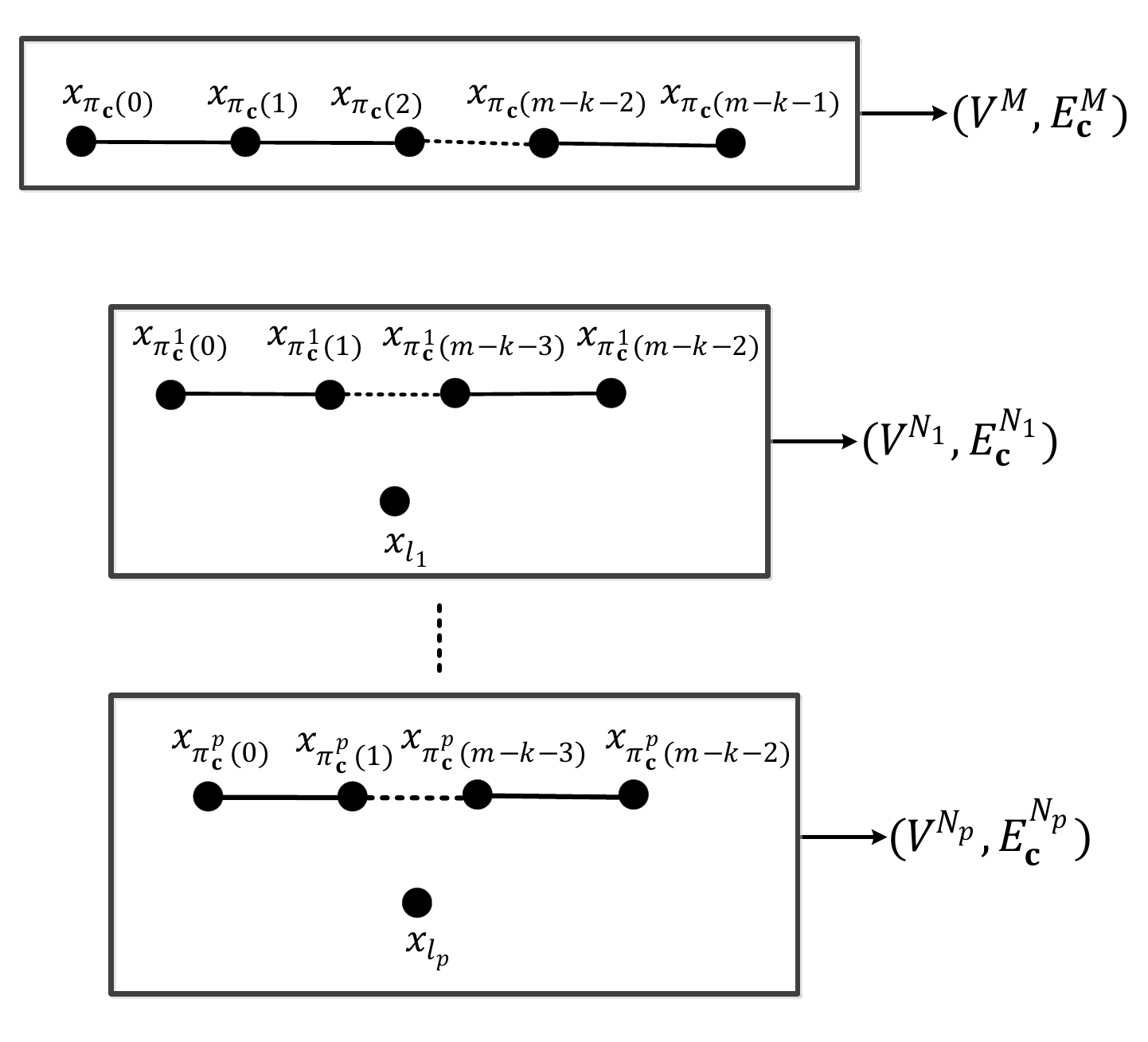}
\caption{The graphs of the restricted Boolean functions corresponding to GBF given in (\ref{cgbf}).}
\end{figure}
As a generalization, in Fig. 3, $(V^M,E_{\textbf{c}}^M)=G(f\arrowvert_{\mathbf{x}_J=\textbf{c}})$, where $f$ is a GBF given in (\ref{cgbf}), $\textbf{c}\in S_M$,
$V_M=\{x_0,x_1,\hdots,x_{m-1}\}\setminus \{x_{j_0},x_{j_1},\hdots,x_{j_{k-1}}\}$, and
$E_{\textbf{c}}^M=\{x_{\pi_\textbf{c}(i)}x_{\pi_\textbf{c}(i+1)}:i=0,1,\hdots,m-k-2\}$. For any two distinct $\textbf{c}_1$, $\textbf{c}_2\in S_M$, the
graphs $(V^M,E_{\textbf{c}_1}^M)$ ($= \!\!G(f\arrowvert_{\mathbf{x}_J=\textbf{c}_1})$) and $(V^M,E_{\textbf{c}_2}^M)$ ($=\!\!G(f\arrowvert_{\mathbf{x}_J=\textbf{c}_2})$)
will be the same if the permutations $\pi_{\textbf{c}_1}$ and $\pi_{\textbf{c}_2}$ are equal.
Otherwise, $E_{\textbf{c}_1}^M\neq E_{\textbf{c}_2}^M$, and hence $(V^M,E_{\textbf{c}_1}^M)$, $(V^M,E_{\textbf{c}_2}^M)$ represent two different graphs.
Similarly, $(V^{N_\delta},E_{\textbf{c}}^{N_\delta})=G(f\arrowvert_{\mathbf{x}_J=\textbf{c}})$, $\textbf{c}\in S_{N_\delta}$ ($\delta=1,2,\hdots,p$),
$V^{N_\delta}=\{x_0,x_1,\hdots,x_{m-1}\}\setminus \{x_{j_0},x_{j_1},\hdots,x_{j_{k-1}},x_{l_\delta}\}$, and
$E_{\textbf{c}}^{N_\delta}=\{x_{\pi_\textbf{c}^\delta(i)}x_{\pi_\textbf{c}^\delta(i+1)}:i=0,1,\hdots,m-k-3\}$,
where $\pi_{\textbf{c}}^\delta$, $\textbf{c}\in S_{N_\delta}$, $\delta=1,2,\hdots,p$ are defined in (\ref{cgbf}).

If a GBF has the same graphical property as given in Fig. 3 and also satisfies the conditions given in \textit{Corollary} \ref{corr_1}, the sequence corresponding
to the GBF lies in a CS of size $2^{k+1}$ and hence the PMEPR is upper bounded by $2^{k+1}$. Similarly, if a GBF meets the condition given in \textit{Corollary} \ref{corr_2} and also has the same graphical
property as in Fig. 3, the sequence corresponding to the GBF lies in a CS of size $2^{k+2}$ with at most PMEPR $2^{k+2}-2M$.

Now, we define the set of vertices as follows:
$P^M_\textbf{c}=\{x_{\pi_{\textbf{c}}(0)},x_{\pi_{\textbf{c}}(m-k-1)}\}$, $\textbf{c}\in S_M$ and
$I^{N^\delta_{\textbf{c}}}=\{x_{\pi_{\textbf{c}}^\delta(0)},x_{\pi_{\textbf{c}}^\delta(m-k-2)}\}$, $\textbf{c}\in S_{N_\delta}$, $\delta=1,2,\hdots,p$.

Schmidt's construction provides a PMEPR upper bound of $2^{k+p+1}$ for the sequences corresponding to the GBFs which have the following
properties:
\begin{itemize}
 \item The restricted Boolean functions of a GBF have the following graphical properties as given in Fig. 3.
 \item $x_{l_\delta}\in P^M_\textbf{c}~\forall \textbf{c}\in S_M,\delta=1,2,\hdots,p$.
 \item $x_{l_\delta}\in I^{N^{\delta_1}_{\textbf{c}}}~\forall\textbf{c}\in S_{N_{\delta_1}}, \delta_1\in \{1,2,\hdots,p\}\setminus\{\delta\},
 \delta=1,2,\hdots,p$.
\end{itemize}
Otherwise, the PMEPR upper bound provided by Schmidt's construction will be strictly greater than $2^{k+p+1}$.
\begin{table}
\centering
\caption{PMEPR Upper Bound for Different Values of $M$ and $p$, where $M+\displaystyle\sum_{i=1}^pN_i<2^m$.}
\begin{tabular}{|l|l|l|l|l||l||}
\hline
$k$                   & Construction                       & $M$                  & $p$                  & \multicolumn{2}{l|}{PMEPR upper bound} \\ \hline
\multirow{6}{*}{1}  & \multirow{2}{*}{\textit{Corollary} \ref{corr_1}}  & \multirow{2}{*}{0} & \multirow{2}{*}{1} & Proposed          & \cite{Schmid2007}                \\ \cline{5-6}
                    &                               &                    &                    & $4$                 & $8$                  \\ \cline{2-6}
                    & \multirow{4}{*}{\textit{Corollary} \ref{corr_2}}  & \multirow{2}{*}{0} & $1$                  & $8$                 & $8$                \\ \cline{4-6}
                    &                               &                    & $2$                  & $8$                 & $\geq 16$          \\ \cline{3-6}
                    &                               & $1$                  & $1$                  & $6$                 & $\geq 8$           \\ \cline{3-6}
                    &                               & $2$                  & $0$                  & $4$                 & $4$                \\ \hline
\multirow{14}{*}{2} & \multirow{3}{*}{\textit{Corollary} \ref{corr_1}}    & \multirow{2}{*}{0}    & $1$                  & $8$                 & $16$ \\ \cline{4-6}
                    &                               &                      &$2$                 & $8$                 & $\geq 32$\\ \cline{3-6}
                    &                               &   $1$                &$1$                 &$8$                  & $\geq 16$\\ \cline{2-6}
                    & \multirow{11}{*}{\textit{Corollary} \ref{corr_2}} & \multirow{4}{*}{0} & $1$                  & $16$                & $16$               \\ \cline{4-6}
                    &                               &                    & $2$                  & $16$                & $\geq 32$          \\ \cline{4-6}
                    &                               &                    & $3$                  & $16$                & $\geq 64$          \\ \cline{4-6}
                    &                               &                    & $4$                  & $16$                & $\geq 128$         \\ \cline{3-6}
                    &                               & \multirow{3}{*}{1} & $1$                  & $14$                & $\geq 16$          \\ \cline{4-6}
                    &                               &                    & $2$                  & $14$                & $\geq 32$          \\ \cline{4-6}
                    &                               &                    & $3$                  & $14$                & $\geq 128$         \\ \cline{3-6}
                    &                               & \multirow{2}{*}{2} & $1$                  & $12$                & $\geq 16$          \\ \cline{4-6}
                    &                               &                    & $2$                  & $12$                & $\geq 32$          \\ \cline{3-6}
                    &                               & $3$                  & $1$                  &$10$                & $\geq 16$          \\ \cline{3-6}
                    &                               & $4$                  & $0$                  & $8$                 & $8$                \\ \hline
\end{tabular}
\end{table}
For different values of $M$ and $p$, we compare the PMEPR upper bounds obtained
from \textit{Corollary} \ref{corr_1} and \textit{Corollary} \ref{corr_2}, with \cite{Schmid2007} in TABLE I.
\section{Code-Rate Comparison with existing work}
In this section, we compare our proposed construction with the constructions given in \cite{pater2000} and \cite{Schmid2007} in terms of code-rate and
PMEPR.
\subsection{Comparison With \cite{pater2000}}
In this subsection, we give a comparison of our proposed construction with \cite{pater2000} to show that the proposed construction can generate more
sequences, i.e., higher code-rate with tighter PMEPR upper bound.

It is observed that by using \textit{Corollary \ref{corr_1}},
we get at least $$\frac{m!}{2}\left[\frac{(m-2)!}{2}-1\right]q^{2m-3}(q-1)^2,$$ complementary sequences with PMEPR at most $4$ and
$$\frac{3m!}{4}\left[\frac{(m-3)!}{2}-1\right]q^{3m-8}(q-1)^2,$$ complementary sequences with PMEPR at most $8$ of length $2^m$. The
detailed derivations on enumeration of complementary sequences with maximum PMEPR $4$ and $8$ are given in
Subsections A and B of Appendix B, respectively.

By \textit{Corollary \ref{corr_2}}, we obtain at least $$\left[\frac{m!(m-2)!(m-1)}{4}\right]q^{2m-2}(q-1)^2,$$ complementary sequences with PMEPR at most $6$
and at least $$m(m-2)\left[\frac{(m-2)!}{2}\right]^2q^{2m-3}(q-1)^2,$$ complementary sequences with PMEPR at most $8$. The
detailed derivations on enumeration of complementary sequences with maximum PMEPR $6$ and $8$ are given in Subsections C and D of Appendix B, respectively.

Now we define three codebooks $\mathcal{S}_1, \mathcal{S}_2, \mathcal{S}_3$ where $\mathcal{S}_1, \mathcal{S}_2$, and $ \mathcal{S}_3$ contain codewords
of length $2^m$ over $\mathbb{Z}_q$ with PMEPR at most $4,6$, and $8$ respectively, given in TABLE II.
\begin{table}[h!]
\tabcolsep=0.1cm
\centering
\caption{PMEPRs and enumerations for codebooks $\mathcal{S}_1, \mathcal{S}_2, \mathcal{S}_3$}
\begin{tabular}{|c|| c| c|}
 \hline
 Codebook & \makecell{PMEPR \\upper bound} & Size of Codebook\\ [0.5ex]
 \hline\hline
 $\mathcal{S}_1$ & 4 & $\frac{m!}{2}\left[\frac{(m-2)!}{2}-1\right]q^{2m-3}(q-1)^2$\\[0.5ex]
 \hline
 $\mathcal{S}_2$ & 6 & $\left[\frac{m!(m-2)!(m-1)}{4}\right]q^{2m-2}(q-1)^2$\\[0.5ex]
 \hline
 $\mathcal{S}_3$ & 8 & $\makecell{\frac{3m!}{4}\left[\frac{(m-3)!}{2}-1\right]q^{3m-8}(q-1)^2\\+m(m-2)\left[\frac{(m-2)!}{2}\right]^2q^{2m-3}(q-1)^2}$\\[0.5ex]
 \hline
\end{tabular}
\end{table}
The code-rate \cite{WCP} of a code-keying OFDM is defined as $\mathcal{R(C)}:=\frac{\log_q |\mathcal{C}|}{L}$, where $\mathcal{|C|}$ and $L$ denote the set size
of codebook $\mathcal{C}$ and the number of subcarriers respectively. In TABLE III and TABLE V, code-rate comparisons with \cite{pater2000} is given. TABLE IV contains code-rates for binary and quaternary cases with
PMEPR at most $6$.
\begin{table}[h!]
\centering
\caption{Code-rate comparison with codebook in \cite{pater2000} with PMEPR at most $4$ Over $\mathbb{Z}_q$}
\begin{tabular}{|l||*{5}{c|}}\hline
\diagbox[dir=NW]{$L=2^m$}{$\mathbb{Z}_q$}
&\multicolumn{4}{c|}{\thead{$q=2$}\vline\thead{$q=4$}}\\ \cline{2-5}
&{\thead{Proposed}}&{\thead{\cite{pater2000}}} & {\thead{Proposed}}&{\thead{\cite{pater2000}}}\\\hline\hline
$m=5$ &$0.4346$&$0.3440$&$0.3762$&$0.1875$\\\hline
$m=6$ &$0.3274$&$0.2660$&$0.2588$&$0.1210$\\\hline
$m=7$ &$ 0.2202$&$0.1800$&$0.1654$&$0.0740$\\\hline
$m=8$ &$0.1398$&$0.1130$&$0.1015$&$0.0440$\\\hline
$m=9$ &$0.0855$&$0.0660$&$0.0605$&$0.0255$\\\hline
$m=10$ &$0.0509$&$0.0380$&$0.0353$&$0.0145$\\\hline
\end{tabular}
\end{table}
\begin{table}[h!]
\centering
\caption{Code-rate for OFDM codes with PMEPR at most $6$ over $\mathbb{Z}_q$}
\begin{tabular}{|l||*{3}{c|}}\hline
\diagbox{$L=2^m$}{$\mathbb{Z}_q$}
&\makebox[8em]{$q=2$}&\makebox[8em]{$q=4$}\\\hline\hline
$m=4$ &$0.6981$&$0.6356$\\\hline
$m=5$ &$0.5466$&$0.4478$\\\hline
$m=6$ &$0.3812$&$0.2935$\\\hline
$m=7$ &$0.2483$&$0.1834$\\\hline
$m=8$ &$0.1547 $&$0.1108$\\\hline
$m=9$ &$0.0933 $&$0.0654$\\\hline
$m=10$ &$0.0549$&$0.0378$\\\hline
\end{tabular}
\end{table}
\begin{table}[h!]
\tabcolsep=0.6cm
\centering
\caption{Code-rate comparison with codebook in \cite{pater2000} with PMEPR at Most $8$ Over $\mathbb{Z}_2$}
\begin{tabular}{|l||*{3}{c|}}\hline
\diagbox{$L=2^m$}{$\mathbb{Z}_q$}
&\multicolumn{2}{c|}{\thead{$q=2$}} \\ \cline{2-3}
&{\thead{Proposed}}&{\thead{\cite{pater2000}}}\\\hline\hline
$m=7$ &$0.2371$&$0.1720$\\\hline
$m=8$ &$0.1501$&$0.1170$\\\hline
$m=9$ &$0.0916$&$0.072$\\\hline
$m=10$ &$0.0544$&$0.043$\\\hline
\end{tabular}
\end{table}
\subsection{Comparison With \cite{Schmid2007}}
In this subsection, we present a comparison between our proposed construction with \cite{Schmid2007} to show that the proposed construction
can provide higher code-rate and PMEPR upper bound.
For $0\leq k<m$, $0\leq r\leq k+1$, and $h\geq 1$, a linear code $\mathcal{A}(k,r,m,h)$ \cite{Schmid2007} is defined to be the set of codewords
corresponding to the set of polynomials
\begin{equation}
 \begin{split}
  \left \{ \sum_{i=0}^{m-k-1}x_\alpha g_i(x_{m-k},\hdots,x_{m-1})+g(x_{m-k},\hdots,x_{m-1}):\right.\\ \left. g_0,\hdots,g_{m-k-1}\in
  \mathcal{F}(r-1,k,h),g\in\mathcal{F}(r,k,h)\right\}.
 \end{split}
\end{equation}
The number of codewords in $\mathcal{A}(k,r,m,h)$ is equal to $2^{s_k}$, where
$$s_k=(m-k)\log_2|\mathcal{F}(r-1,k,h)|+\log_2|\mathcal{F}(r,k,h)|.$$
Now, $\mathcal{R}(k,m,h)$ \cite{Schmid2007} is defined to be the set of codewords associated with the following polynomials over $\mathbb{Z}_{2^h}$
\begin{equation}
\begin{split}
 2^{h-1}\!\!\!\!\!\!\!\sum_{\textbf{c}\in \{0,1\}^k}\!\!\!\!\!\sum_{i=0}^{m-k-2}\!\!\!\!\!x_{\pi_{\textbf{c}}(i)}
 x_{\pi_{\textbf{c}}(i+1)}\!\!\prod_{j=0}^{k-1}\!\!x_{m-k+j}^{c_j}(1-x_{m-k+j})^{(1-c_j)},
 \end{split}
\end{equation}
where $\pi_{\textbf{c}}$ are $2^k$ permutations of $\{0,1,\hdots,m-k-1\}$. For $m-k>1$ and $r>2-h$, the set $\mathcal{R}(k,m,h)$ contains
$[ (m-k)!/2]^{2^{\min\{r+h-3,k\}}}$ codewords corresponding to a GBF of effective-degree at most $r$. The sequences in the cosets of
$\mathcal{A}(k,r,m,h)$ with coset representatives in $\mathcal{R}(k,m,h)$ have PMEPR at most $2^{k+1}$ and the code has minimum Lee and squared Euclidean
distance equal to $2^{m-r}$ and $2^{m-r+2}\sin^2(\frac{\pi}{2^h})$ respectively. We define $I^m_k=\{0,1,\hdots,m-k-1\}$ which will be
used in the construction of code.
\subsubsection{Code Construction by Using \textit{Corollary} \ref{corr_1}}
In this section, we consider the case $p=1$, $M=0$ of \textit{Corollary} \ref{corr_1}. For $0\leq k<m-1$, $0\leq r\leq k+1$, $\alpha\neq l_1$ ($l_1\in\{0,1,\hdots,m-k-1\}$) and $h\geq 1$, we
define a linear code $\mathcal{A}_{1,l_1}(k,r,m,h)$ corresponding to the set of polynomials
\begin{equation}
 \begin{split}
  \left \{ \sum_{i=0}^{m-k-1}x_\alpha g_i(x_{m-k},\hdots,x_{m-1})+g(x_{m-k},\hdots,x_{m-1}):\right.\\ \left. g_0,\hdots,g_{m-k-1}\in
  \mathcal{F}(r-1,k,h),g\in\mathcal{F}(r,k,h)\right \}.
 \end{split}
\end{equation}
$\mathcal{A}_{1,l_1}(k,r,m,h)$ contains $2^{s_{1,k}}$ codewords, where $$s_{1,k}=(m-k-1)\log_2|\mathcal{F}(r-1,k,h)|+\log_2|\mathcal{F}(r,k,h)|.$$ Since,
$\mathcal{A}_{1,l_1}(k,r,m,h)\subset \mathcal{A}(k,r,m,h)$, the minimum distances of $\mathcal{A}_{1,l_1}(k,r,m,h)$ can be lower bounded by $2^{m-r}$ and $2^{m-r+2}\sin^2(\frac{\pi}{2^h})$.

Now, we assume that $\mathcal{R}_{1,l_1}(k,m,h)$ be the set of codewords associated with the following polynomials
\begin{equation}
\begin{split}
 2^{h-1}\!\!\!\!\!\!\!\sum_{\textbf{c}\in \{0,1\}^k}\!\!\!\!\!\sum_{i=0}^{m-k-3}\!\!\!\!\!x_{\pi_{\textbf{c}}(i)}
 x_{\pi_{\textbf{c}}(i+1)}\!\!\prod_{j=0}^{k-1}\!\!x_{m-k+j}^{c_j}(1-x_{m-k+j})^{(1-c_j)}\\+2^{h-1}x_{l_1}(e_0x_{m-1}+\cdots+e_{k-1}x_{m-k}),\qquad\qquad
 \end{split}
\end{equation}
where $\pi_{\textbf{c}}$ are $2^k$ permutations of $\{0,1,\hdots,m-k-1\}\setminus {l_1}$ and $e_0,\hdots,e_{k-1}\in \{0,1\}$, but all can not be zero at
the same time.

For $m-k>2$ and $r>2-h$, it can be shown that the set $\mathcal{R}_{1,l_1}(k,m,h)$ contains $(2^k-1)[(m-k-1)!/2]^{2^{\min(r+h-3,k)}}$ codewords corresponding to a GBF
of effective degree at most $r$.
\begin{note}\label{note_1}
Assume that $m-k>2$. Let $2\leq r\leq k+2$ when $h=1$, $1\leq r\leq k+1$ when $h>1$ and $r'=\min\{r,k+1\}$.
By using \textit{Corollary} \ref{corr_1}, it can be shown that any coset of $\mathcal{A}_{1,l_1}(k,r',m,h)$ with coset representatives
in $\mathcal{R}_{1,l_1}(k,m,h)$ have PMEPR at most $2^{k+1}$.
Now take the union of $(2^k-1)[(m-k-1)!/2]^{2^{\min(r+h-3,k)}}$ distinct cosets of $\mathcal{A}_{1,l_1}(k,r',m,h)$, each containing a word
in $\mathcal{R}_{1,l_1}(k,m,h)$ with effective degree at most $r$. The PMEPR of the corresponding polyphase codewords in this code is at most $2^{k+1}$.
Since the code is a subcode
of $\textnormal{ERM}(r,m,h)$, its minimum Lee and squared Euclidean distances are lower bounded by $2^{m-r}$ and $2^{m-r+2}\sin^2(\frac{\pi}{2^h})$ respectively.
\end{note}
\subsubsection{Code Construction by Using \textit{Corollary} \ref{corr_2}}
In this section, we consider the case $p\geq 2$, $M=0$ of \textit{Corollary} \ref{corr_2}. Consider $\mathcal{R}_{2,\textbf{l}}(k,m,h)$ be the set of
codewords associated with the following polynomials
\begin{equation}
\begin{split}
 2^{h-1}\sum_{\alpha=1}^p\!\sum_{\textbf{c}_\alpha\in S_{N_{\alpha}}}\sum_{i=0}^{m-k-3}x_{\pi_{\textbf{c}_\alpha}(i)}
 x_{\pi_{\textbf{c}_\alpha}(i+1)}\\\times\prod_{j=0}^{k-1}x_{m-k+j}^{c^{\alpha}_j}(1-x_{m-k+j})^{(1-c^{\alpha}_j)},
 \end{split}
\end{equation}
where $\textbf{c}_\alpha=(c^\alpha_0,\hdots,c^\alpha_{k-1})$, $\pi_{\textbf{c}_\alpha}$ are $N_\alpha$ permutations of
$\{0,1,\hdots,m-k-1\}\setminus l_\alpha$, $\textbf{l}=(l_1,l_2,\hdots,l_p)$ and $\sum_{\alpha=1}^p N_\alpha=2^k$.

Now, we define the set $\mathcal{L}=\left\{\textbf{l}:\textbf{l}\in \{0,1,\hdots,m-k-1\}^p, l_1<l_2<\cdots<l_p\right\}$.

For $m-k>2$, $r>2-h$, and $\textbf{l}\in \mathcal{L}$, it can be shown that the set $\mathcal{R}_{2,\textbf{l}}(k,m,h)$ contains
\begin{equation}\nonumber
\begin{split}
   [(m\!\!-\!\!k\!\!-\!\!1)!/2]^{\min(2^{r+h-3},N_1)}\times[(m\!\!-\!\!k\!\!-\!\!1)!/2]^{\min(2^{r+h-3},N_2)}\\\times \cdots\times [(m\!\!-\!\!k\!\!-\!\!1)!/2]^{\min(2^{r+h-3},N_p)}
\end{split}
\end{equation}
 codewords corresponding to a GBF
of effective degree at most $r$.
\begin{note}\label{note_2}
Assume $m-k>2$. Let $2\leq r\leq k+2$ when $h=1$, $1\leq r\leq k+1$ when $h>1$ and $r'=\min\{r,k+1\}$.
 By using \textit{Corollary} \ref{corr_2}, it can be shown that any coset of $\mathcal{A}(k,r',m,h)$  with coset representatives
 in $\mathcal{R}_{2,\textbf{l}}(k,m,h)$ have at most PMEPR $2^{k+2}$. It is also observed that the minimum Lee and squared Euclidean distances of the code
 $$\displaystyle\bigcup_{\textbf{a}\in \mathcal{R}_{2,\textbf{l}}(k,m,h)}\left(\textbf{a}+\mathcal{A}(k,r,m,h)\right)$$ are lower bounded
 by $2^{m-r}$ and $2^{m-r+2}\sin^2(\frac{\pi}{2^h})$ respectively.
\end{note}
\subsubsection{Code Construction With Maximum PMEPR $4$ and $8$}
In this part, we construct codes with maximum PMEPR $4$ and $8$ by using the above discussed codes.
\begin{corollary}[Code With Maximum PMEPR $4$]\label{corr_4}
Assume that $m>3$. Let $2\leq r\leq 3$ when $h=1$, $1\leq r\leq 2$ when $h>1$ and $r'=\min\{r,2\}$. Now, consider
\begin{equation}
 \begin{split}
  \mathcal{C}=&\left[\bigcup_{\textbf{a}_1\in \mathcal{R}(1,m,h)}\!\!\!\!\!\!\!\textbf{a}_1\!\!+\!\!
  \mathcal{A}(1,r',m,h)\right]\!\!\\&\bigcup\!\!
  \left[\bigcup_{l_1\in I^{m}_1}\left(\bigcup_{\textbf{a}_2\in \mathcal{R}_{1,l_1}(1,m,h)}\!\!\!\!\!\!\!\!\!\!\!\textbf{a}_2\!\!+\!\!\mathcal{A}_{1,l_1}(1,r',m,h)\right)\right].
 \end{split}
\end{equation}

The code $|\mathcal{C}|$ contains codewords or sequences with at most PMEPR $4$. Hence, the maximum PMEPR of $\mathcal{C}$ is $4$.
We denote the number of codewords or sequences in the code by $|\mathcal{C}|$, where
\begin{equation}\label{pmepr4}
\begin{split}
 |C|=&\left(2^{s_1}\times[ (m-1)!/2]^{2^{\min\{r+h-3,1\}}}\right)\\&+\left(2^{s_{1,1}}\times(m-1)\times[(m-2)!/2]^{2^{\min(r+h-3,1)}}\right).
 \end{split}
\end{equation}
Since $\mathcal{C}$ is a subcode of $\textnormal{ERM}(r,m,h)$, the minimum Lee and squared Euclidean distances of the code are lower bounded by
$2^{m-r}$ and $2^{m-r+2}\sin^2(\frac{\pi}{2^h})$ respectively.

\begin{table}[h!]
\centering
\caption{Code-rate comparison with codebook in \cite{Schmid2007} with maximum PMEPR $4$ Over $\mathbb{Z}_{2^h}$}
\begin{tabular}{ |c|c|c|c|c|c|c| }
\hline
$m$ & $h$ & $r$ &Proposed&\cite{Schmid2007}& $d_L$ & $d_E^2$ \\
\hline
\hline
$4$ &$1$ &$2$ &$0.6192$ &$0.5990$ &$4$ &$16.00$\\

&        &$3$ &$0.7053$ &$0.6980$ &$2$ &$8.00$\\

    &$2$ &$1$ &$0.4611$ &$0.4560$ &$8$ &$16.00$\\

&        &$2$ &$0.6000$  &$0.5990$ &$4$ &$8.00$\\
\hline
$5$ &$1$ &$2$ &$0.4345$ &$0.4250$ &$8$ &$32.00$\\

&        &$3$ &$0.5392$ &$0.5366$ &$4$ &$16.00$\\

    &$2$ &$1$ &$0.3087$ &$0.3060$ &$16$ &$32.00$\\

&        &$2$ &$0.4249$ &$0.4246$ &$8$  &$16.00$\\
\hline
$6$ &$1$ &$2$ &$0.2848$ &$0.2798$ &$16$ &$64.00$\\

&        &$3$ &$0.3732$ &$0.3721$ &$8$  &$32.00$\\

    &$2$ &$1$ &$0.1959$ &$0.1946$ &$32$ &$64.00$\\

&        &$2$ &$0.2799$ &$0.2798$ &$16$ &$32.00$\\
\hline
\end{tabular}
\end{table}
From (\ref{pmepr4}), it is clear that the set size of the sequences with maximum PMEPR $4$ obtained from our proposed construction is larger than the set size
given in \cite{Schmid2007}.
In TABLE VI, we have compared the code-rate of sequences with maximum PMEPR $4$ obtained from our proposed construction with that of the construction given
in \cite{Schmid2007}.
\end{corollary}
\begin{corollary}[Code With Maximum PMEPR $8$]\label{corr_5}
 Suppose $m>4$. Let $2\leq r\leq 4$ when $h=1$, $1\leq r\leq 3$ when $h>1$ and $r''=\min\{r,3\}$.

 For the case $2\leq r\leq 3$ when $h=1$, $1\leq r\leq 2$ when $h>1$ and $r''=\min\{r,3\}$, we consider the code $\mathcal{C}_1$, defined by
 \begin{equation}
  \begin{split}
   \mathcal{C}_1=&\left[\bigcup_{\textbf{b}_1\in \mathcal{R}(2,m,h)}\!\!\!\!\!\!\!\!\textbf{b}_1\!\!+\!\!\mathcal{A}(2,r'',m,h)\!\!\right]\!\!\\&\bigcup\!\!
  \left[\bigcup_{l_1\in I^m_2}\left(\bigcup_{\textbf{b}_2\in \mathcal{R}_{1,l_1}(2,m,h)}\!\!\!\!\!\!\!\!\!\textbf{b}_2\!\!+\!\!\mathcal{A}_{1,l_1}(2,r'',m,h)\right)\!\!\right]\\
  &\bigcup\left[\bigcup_{\textbf{l}\in \mathcal{L}}\left(\bigcup_{\textbf{b}_3\in \mathcal{R}_{2,\textbf{l}}(1,m,h)}\!\!\!\!\!\!\!\!\!\textbf{b}_3\!\!+\!\!\mathcal{A}(1,r',m,h)\right)\!\!\right],
  \end{split}
 \end{equation}
 where
 \begin{equation}\label{pmepr81}
  \begin{split}
   |\mathcal{C}_1|=&\left(2^{s_2}\times[ (m-2)!/2]^{2^{\min\{r+h-3,2\}}}\right)\\&
   +\left(3\times(m-2)\times 2^{s_{1,2}}\times[(m-3)!/2]^{2^{\min(r+h-3,2)}}\right)\\&
   +\left(2^{s_1}\times|\mathcal{L}|\times[ (m-2)!/2]^{2\times \min\{2^{r+h-3},1\}} \right),
  \end{split}
 \end{equation}
 where $|\mathcal{L}|={m-1\choose 2}$ for $k=1$ and $p=2$.

Since $\mathcal{C}_1$  is a subcode of $\textnormal{ERM}(r,m,h)$, the minimum Lee and squared Euclidean distances of the code are lower bounded by
$2^{m-r}$ and $2^{m-r+2}\sin^2(\frac{\pi}{2^h})$ respectively.

For $r=4$ when $h=1$ and $r=3$ when $h>1$, we consider the code $\mathcal{C}_2$, defined by
\begin{equation}
  \begin{split}
   \mathcal{C}_2=&\left[\bigcup_{\textbf{b}_1\in \mathcal{R}(2,m,h)}\!\!\!\!\!\!\!\!\!\!\textbf{b}_1\!\!+\!\!\mathcal{A}(2,r'',m,h)\!\!\right]\\&\bigcup\!\!
  \left[\bigcup_{l_1\in I^m_2 }\left(\bigcup_{\textbf{b}_2\in \mathcal{R}_{1,l_1}(2,m,h)}\!\!\!\!\!\!\!\!\!\!\!\!\textbf{b}_2\!\!+\!\!\mathcal{A}_{1,l_1}(2,r'',m,h)\right)\!\!\right],
  \end{split}
 \end{equation}
 where
\begin{equation}\label{pmepr82}
  \begin{split}
   |\mathcal{C}_2|&=\left(2^{s_2}\times[ (m-2)!/2]^{2^{\min\{r+h-3,2\}}}\right)\\&
   +\left(3\times 2^{s_{1,2}}\times (m-2)\times[(m-3)!/2]^{2^{\min(r+h-3,2)}}\right).
  \end{split}
 \end{equation}
 Since $\mathcal{C}_2$ is a subcode of $\textnormal{ERM}(r,m,h)$, the minimum Lee and squared Euclidean distances of the code are lower bounded by
$2^{m-r}$ and $2^{m-r+2}\sin^2(\frac{\pi}{2^h})$ respectively.
\begin{table}[h!]
\centering
\caption{Code-rate comparison with codebook in \cite{Schmid2007} with maximum PMEPR $8$ Over $\mathbb{Z}_{2^h}$}
\begin{tabular}{ |c|c|c|c|c|c|c| }
\hline
$m$ & $h$ & $r$ &Proposed&\cite{Schmid2007}& $d_L$ & $d_E^2$ \\
\hline
\hline
$5$ &$1$ &$2$ &$0.5138$ &$0.4558$ &$8$ &$32.00$\\

&        &$3$ &$0.6056$ &$0.5991$ &$4$  &$16.00$\\

&        &$4$ &$0.6984$ &$0.6981$ &$2$  &$8.00$\\

    &$2$ &$1$ &$0.3495$ &$0.3216$ &$16$ &$32.00$\\

&        &$2$ &$0.5030$ &$0.5025$ &$8$  &$16.00$\\

&        &$3$ &$0.5991$ &$0.5991$ &$4$  &$8.00$\\
\hline
$6$ &$1$ &$2$ &$0.3552$ &$0.3060$ &$16$ &$64.00$\\

&        &$3$ &$0.4263$ &$0.4245$ &$8$  &$32.00$\\

&        &$4$ &$0.5366$ &$0.5366$ &$4$  &$16.00$\\

    &$2$ &$1$ &$0.2322$ &$0.2077$ &$32$ &$64.00$\\
&        &$2$ &$0.3374$ &$0.3372$ &$16$ &$32.00$\\
&        &$3$ &$0.4246$ &$0.4246$ &$8$  &$16.00$\\
\hline
\end{tabular}
\end{table}

From (\ref{pmepr81}) and (\ref{pmepr82}), it is clear that our proposed construction can provide more number of sequences than the construction given in \cite{Schmid2007}.
In TABLE VII, we have compared the code-rate of sequences with maximum PMEPR $8$ obtained from our proposed construction with that of the construction given
in \cite{Schmid2007}.
\end{corollary}
\subsection{Comparison with \cite{liumc,imli,liug,avi_commu,sdas,Sdas_lett,pskaccess,psktcom,chenzcom,chentcom,chentit,chencommlett,chenconf,ara_bzcp_2018,liuqcss,liano}}
In this subsection, we give a comparison of our proposed construction with the works introduced
in \cite{liumc,imli,liug,avi_commu,sdas,Sdas_lett,pskaccess,psktcom,chenzcom,chentcom,chentit,chencommlett,chenconf,ara_bzcp_2018,liuqcss,liano}. The comparison has been given in TABLE VIII.
\begin{table*}[h!]\small
\centering
\caption{Comparison with \cite{liumc,imli,liug,avi_commu,sdas,Sdas_lett,pskaccess,psktcom,chenzcom,chentcom,chentit,chencommlett,chenconf,ara_bzcp_2018,liuqcss,liano}}
\resizebox{\textwidth}{!}{
\begin{tabular}{ |c|c|c|c|c| }
\hline
Sequence Class                        & Approach &Phase& Length&Constraints \\\hline\hline
Complete complementary codes (CCC) \cite{liumc}&Second-order GBFs&$q$&$2^m$& $m\geq 1,q\geq2,2|q$\\\hline
General QAM Golay complementary seq. \cite{imli}&PSK GDJ seq.&$q$& $2^m$&$m\geq 1$\\\hline
General QAM Golay complementary seq. \cite{liug}&Gaussian integer pairs&$q$&$2^m$&$m\geq 1$\\\hline
CS \cite{avi_commu} &Seq. Insertion&$q$&$N+1,N+2,2N+3$&$q\geq2,2|q$, $N$ length of a GCP  \\\hline
CCC \cite{sdas} &Paraunitary matrices&$q$&$M^{N'}$&$M>1$, $N'\geq 1$, $q\geq 2$\\\hline
CCC \cite{Sdas_lett} &Paraunitary matrices&$q$& $P^{N'}$&$P|M,N'\geq 1,q\geq 2$\\\hline
Inter-group complementary code set \cite{pskaccess} &Second-order GBFs&$q$&$2^m$&$m\geq 2,q\geq 2,2|q$\\\hline
Z-complementary code set \cite{psktcom} &Second-order GBFs&$q$&$2^m$&$m\geq 2,q\geq 2,2|q$\\\hline
Z-complementary code set \cite{chenzcom} &Second-order GBFs&$q$&$2^m$&$m\geq 3,q\geq 2,2|q$\\\hline
CS with large zero-correlation zone \cite{chentcom} &Second-order GBFs&$q$&$2^m$&$m\geq 2,q\geq 2,2|q$\\\hline
CS \cite{chentit} &Second-order GBFs&$q$&$2^{m-1}+2^v$ &$m\geq 2,1\leq v\leq m-1,q\geq 2,2|q$\\\hline
CCC \cite{chencommlett} &RM codes&$q$&$2^m$&$m\geq 2,q\geq 2,2|q$\\\hline
CS \cite{chenconf} &RM codes&$q$&$2^m$&$m\geq 2, q\geq 2,2|q$\\\hline
Z-complementary pair \cite{ara_bzcp_2018}&Seq. Insertion and concatenation&$q$&$2^{\alpha+2}{10}^\beta{26}^\gamma$&$\alpha,\beta,\gamma\geq 0,q=2$\\\hline
Quasi-complementary seq. set (QCSS) \cite{liuqcss} &Singer difference sets and optimal quaternary seq. set &$q$&$2^m-1,2(2^m-1)$&$q=2^m-1,m\geq 2$\\\hline
QCSS \cite{liano} & Primitive elements of extension field and trace function &$q$&$q$, $q-1$&$q\geq 3,q=p^n,n\geq 1$, $p$ prime\\\hline
\textit{Corollary 1}&GBFs of order no less than $2$&$q$&$2^m$&$m\geq 2, q\geq 2,2|q$\\\hline
\textit{Corollary 2}&GBFs of order no less than $2$&$q$&$2^m$&$m\geq 2, q\geq 2,2|q$\\\hline
\end{tabular}}
\end{table*}
\section{Conclusions}
In this paper, we proposed a direct and generalized construction of polyphase CS by using higher order GBFs and the concept of isolated vertices.
The proposed construction provides tighter PMEPR upper bound for code words and higher code-rate by maintaining the same minimum code
distances as compared to Schmidt's construction.
We have shown that our proposed construction gives rise to sequences with maximum
PMEPR upper bound of $4$ in \textit{Corollary} \ref{corr_1} and $8$ in both \textit{Corollary} \ref{corr_1} and \textit{Corollary} \ref{corr_2}, respectively. In addition, we have obtained sequences with maximum PMEPR upper
bound of $6$ in \textit{Corollary} \ref{corr_2}.
The constructions given by Davis and Jedwab \cite{Davis1999}, Paterson \cite{pater2000} and Schmidt \cite{Schmid2007}
appear as special cases of our proposed construction. Lastly, as pointed out by one reviewer, the practical PMEPR performances of our
constructed sequences also depend on the power amplifier (PA) at the transmitter. The PA may introduce certain non-linear distortions
when the transmit signals are not in the linear amplification zone. As a future work of this research, it would be interesting 1) to evaluate
the reduction of the input back-off (IBO) for different PAs based on our constructed sequences and compare it with the known sequences.
2) to compare the complementary commulative distribution function (CCDF) of the PMPER of our proposed method to the known methods.
\begin{appendices}
\section{Proof of \textnormal{\textit{Theorem} \ref{Theorem1}}}
For any $\tau\neq 0$, the sum of AACF of sequences from the set $S$, which is defined in (\ref{defis}), can be written as
\begin{equation}\label{A=L_1+L_2}
\displaystyle \sum_{\textbf{d}d''}A\left(f+\frac{q}{2}(\mathbf{d}\cdot\mathbf{x}_J+d''e_2)\right)(\tau)=\mathcal{L}_1+\mathcal{L}_2,
\end{equation}
where
\begin{equation}\label{L_1}
\mathcal{L}_1=\displaystyle \sum_{\textbf{d}d''}\displaystyle \sum_{\textbf{c}}A\left(\left(f+\frac{q}{2}(\mathbf{d}\cdot\mathbf{x}_J+d''e_2)
\right)\big\arrowvert_{\textbf{x}_J=\textbf{c}}\right)(\tau),
\end{equation}
and
\begin{equation}\label{L_2}
\begin{split}
\mathcal{L}_2 =&\displaystyle \sum_{\textbf{d}d''}\displaystyle \sum_{\textbf{c}_1\neq \textbf{c}_2}C \left( \left(f+\frac{q}{2}(\mathbf{d}\cdot\mathbf{x}_J+d''e_2)\right)\big\arrowvert_{\textbf{x}_J=\textbf{c}_1},
\right.\\&\left.~~~~~~~~~~~~~\left(f+\frac{q}{2}(\mathbf{d}\cdot\mathbf{x}_J+d''e_2)\right)\big\arrowvert_{\textbf{x}_J=\textbf{c}_2} )\right) (\tau).
\end{split}
\end{equation}
We first focus on the term $\mathcal{L}_1$, which can be written as
\begin{equation}\label{L_1=T+T_i}
 \mathcal{L}_1=T+\sum_{i=1}^pT_i,
\end{equation}
where
\begin{equation}\label{T}
 T=\displaystyle \sum_{\textbf{d}d''}\displaystyle \sum_{\textbf{c}\in S_M}A\left(\left(f+\frac{q}{2}(\mathbf{d}\cdot\mathbf{x}_J+d''e_2)
\right)\big\arrowvert_{\textbf{x}_J=\textbf{c}}\right)(\tau),
\end{equation}
and
\begin{equation}\label{T_i}
T_i=\displaystyle \sum_{\textbf{d}d''}\displaystyle \sum_{\textbf{c}\in S_{N_i}}A\left(\left(f+\frac{q}{2}(\mathbf{d}\cdot\mathbf{x}_J+d''e_2)
\right)\big\arrowvert_{\textbf{x}_J=\textbf{c}}\right)(\tau),
\end{equation}
where $S_M$ is the set of all those $\textbf{c}$ for which $G(f\big\arrowvert_{\textbf{x}_J=\textbf{c}})$ is a path over $m-k$ vertices.

To find $\mathcal{L}_1$, we first start with $T$. Since, $G(f\big\arrowvert_{\textbf{x}_J=\textbf{c}})$ is a path over $m-k$ vertices for all $\textbf{c}\in S_M$, we have \cite{pater2000}
\begin{equation}
\begin{split}
\displaystyle \sum_{d''}&A\left(\left(f\!+\!\frac{q}{2}(\mathbf{d}\cdot\mathbf{x}_J\!+\!d''e_2)
\right)\big\arrowvert_{\textbf{x}_J=\textbf{c}}\right)(\tau)\\&=\!\!
\begin{cases}
 2^{m-k+1}, & \tau=0,\\
 0, \!\!\!\!\!&\!\!\!\!\! \textnormal{otherwise.}
\end{cases}
\end{split}
\end{equation}
Therefore,
\begin{equation}\label{Tderi}
 \begin{split}
  T&=\displaystyle \sum_{\textbf{d}d''}\displaystyle \sum_{\textbf{c}\in S_M}A\left(\left(f+\frac{q}{2}(\mathbf{d}\cdot\mathbf{x}_J+d''e_2)
\right)\big\arrowvert_{\textbf{x}_J=\textbf{c}}\right)(\tau)\\&
=\begin{cases}
   2^{m+1}M, & \tau=0,\\
   0, & \textnormal{otherwise.}
  \end{cases}
 \end{split}
\end{equation}
To find $\mathcal{L}_1$, it remains to find $T_i$ $(i=1,2,\hdots,p)$ where
\begin{equation}\nonumber
T_i=\displaystyle \sum_{\textbf{d}d''}\displaystyle \sum_{\textbf{c}\in S_{N_i}}A\left(\left(f+\frac{q}{2}(\mathbf{d}\cdot\mathbf{x}_J+d''e_2)
\right)\big\arrowvert_{\textbf{x}_J=\textbf{c}}\right)(\tau).
\end{equation}
We can express each of $T_i$, as
\begin{equation}\label{T_ider}
\begin{split}
 T_i&=\displaystyle \sum_{\textbf{d}d''}\displaystyle \sum_{\textbf{c}\in S_{N_i}}A\left(\left(f+\frac{q}{2}(\mathbf{d}\cdot\mathbf{x}_J+d''e_2)
\right)\big\arrowvert_{\textbf{x}_J=\textbf{c}}\right)(\tau)\\
&=\!\!\!\displaystyle \sum_{\textbf{d}d''}\displaystyle \sum_{\textbf{c}\in S_{N_i}}\!\!\sum_{\beta\in\{0,1\}}\!\!\!\!\!\!
A\!\!\left(\left(f\!\!+\!\frac{q}{2}(\mathbf{d}\cdot\mathbf{x}_J\!+\!d''e_2)
\right)\big\arrowvert_{\textbf{x}_Jx_{l_i}=\textbf{c}\beta}\right)(\tau)\\
&+\displaystyle \sum_{\textbf{d}d''}\displaystyle \sum_{\textbf{c}\in S_{N_i}}\sum_{\beta\in\{0,1\}}
\!\!\!\!C\left(\left(f\!+\!\frac{q}{2}(\mathbf{d}\cdot\mathbf{x}_J\!+\!d''e_2)
\right)\big\arrowvert_{\textbf{x}_Jx_{l_i}=\textbf{c}\beta},\right. \\&\qquad\qquad\left. \left(f+\frac{q}{2}(\mathbf{d}\cdot\mathbf{x}_J+d''e_2)
\right)\big\arrowvert_{\textbf{x}_Jx_{l_i}=\textbf{c}(1-\beta)}\right)(\tau).
\end{split}
\end{equation}
Since, for all $\textbf{c}\in S_{N_i}$, $G(f\big\arrowvert_{\textbf{x}_J=\textbf{c}})$ consists of a path over $m-k-1$ vertices and
one isolated vertex labeled $l_i$, $G(f\big\arrowvert_{\textbf{x}_Jx_{l_i}=\textbf{c}\beta})$ is a path over $m-k-1$ vertices.
Therefore
\begin{equation}
\begin{split}
 \sum_{d''}A&\left(\left(f+\frac{q}{2}(\mathbf{d}\cdot\mathbf{x}_J+d''e_2)
\right)\big\arrowvert_{\textbf{x}_Jx_{l_i}=\textbf{c}\beta}\right)(\tau)\\&=
\begin{cases}
 2^{m-k}, & \tau=0,\\
 0, & \textnormal{otherwise.}
\end{cases}
\end{split}
\end{equation}
Hence, the first auto-correlation term in (\ref{T_ider}) can be expressed as
\begin{equation}\label{T_iauto}
 \begin{split}
  \displaystyle \sum_{\textbf{d}d''}\displaystyle \sum_{\textbf{c}\in S_{N_i}}\sum_{\beta\in\{0,1\}}
A&\left(\left(f+\frac{q}{2}(\mathbf{d}\cdot\mathbf{x}_J+d''e_2)
\right)\big\arrowvert_{\textbf{x}_Jx_{l_i}=\textbf{c}\beta}\right)(\tau)\\&=
\begin{cases}
 2^{m+1}N_i, & \tau=0,\\
 0, & \textnormal{otherwise.}
\end{cases}
 \end{split}
 \end{equation}
Since, for all $\textbf{c}\in S_{N_i}$, $G(f\big\arrowvert_{\textbf{x}_J=\textbf{c}})$ consists of a path and one isolated vertex  $x_{l_i}$, the only
term involving $x_{l_i}$ is with the variables of the deleted vertices. Thus the only term in $x_{l_i}$ in $f$ can be expressed as follows.
\begin{equation}\label{newdef}
 \begin{split}
 \sum_{r=1}^k\sum_{0\leq i_1<i_2<\cdots<i_r<k}\!\!\!\!\!\!\!\!\!\!\!\!\!\!
\varrho^{l_i}_{i_1,i_2,\hdots, i_r}x_{j_{i_1}}x_{j_{i_2}}\cdots x_{j_{i_r}}x_{l_i}
=L^{l_i}_{\mathbf{x}_J}x_{l_i},
\end{split}
\end{equation}
where
\begin{equation}\nonumber
 \begin{split}
 L^{l_i}_{\mathbf{x}_J}&=\displaystyle\sum_{r=1}^k\sum_{0\leq i_1<i_2<\cdots<i_r<k}\!\!\!\!\!\!\!\!\!\!\!\!\!\!
\varrho^{l_i}_{i_1,i_2,\hdots, i_r}x_{j_{i_1}}x_{j_{i_2}}\cdots x_{j_{i_r}}.
\end{split}
\end{equation}
To simplify the cross-correlation term in (\ref{T_ider}), we have the following equality by \textit{Lemma} \ref{lemmaa} and (\ref{newdef}).
\begin{equation}\nonumber
 \begin{split}
  \sum_{d''}C&\left(\left(f+\frac{q}{2}(\mathbf{d}\cdot\mathbf{x}_J+d''e_2)
\right)\big\arrowvert_{\textbf{x}_Jx_{l_i}=\textbf{c}\beta},\right. \\&\qquad\left. \left(f+\frac{q}{2}(\mathbf{d}\cdot\mathbf{x}_J+d''e_2)
\right)\big\arrowvert_{\textbf{x}_Jx_{l_i}=\textbf{c}(1-\beta)}\right)(\tau)\\&
=\begin{cases}
\omega_q^{(2\beta-1)g_{l_i}}\omega_q^{(2\beta-1)L^{l_i}_{\textbf{c}}}2^{m-k}, & \tau=(2\beta-1)2^{l_i},\\
0, & \textnormal{otherwise},
\end{cases}
\end{split}
\end{equation}
where $\beta\in \{0,1\}$.

Therefore, the cross-correlation term of (\ref{T_ider})  is simplified as
\begin{equation}\label{T_icorr}
 \begin{split}
 \displaystyle \sum_{\textbf{d}d''}&\displaystyle \sum_{\textbf{c}\in S_{N_i}}\sum_{\beta\in\{0,1\}}
C\left(\left(f+\frac{q}{2}(\mathbf{d}\cdot\mathbf{x}_J+d''e_2)
\right)\big\arrowvert_{\textbf{x}_Jx_{l_i}=\textbf{c}\beta},\right. \\& \left. \left(f+\frac{q}{2}(\mathbf{d}\cdot\mathbf{x}_J+d''e_2)
\right)\big\arrowvert_{\textbf{x}_Jx_{l_i}=\textbf{c}(1-\beta)}\right)(\tau)\\
&=\begin{cases}
   \omega_q^{g_{l_i}}2^{m}\displaystyle\sum_{\textbf{c}\in S_{N_i}}\omega_q^{L^{l_i}_{\textbf{c}}}, & \tau=2^{l_i},\\
   \omega_q^{-g_{l_i}}2^{m}\displaystyle\sum_{\textbf{c}\in S_{N_i}}\omega_q^{-L^{l_i}_{\textbf{c}}}, & \tau=-2^{l_i},\\
   0, & \textnormal{otherwise.}
  \end{cases}
 \end{split}
\end{equation}
From (\ref{T_ider}), (\ref{T_iauto}) and (\ref{T_icorr}), we have
\begin{equation}\label{T_iderivation}
 \begin{split}
  T_i=\begin{cases}
       2^{m+1}N_i,& \tau=0,\\
       \omega_q^{g_{l_i}}2^{m}\displaystyle\sum_{\textbf{c}\in S_{N_i}}\omega_q^{L^{l_i}_{\textbf{c}}}, & \tau=2^{l_i},\\
   \omega_q^{-g_{l_i}}2^{m}\displaystyle\sum_{\textbf{c}\in S_{N_i}}\omega_q^{-L^{l_i}_{\textbf{c}}}, & \tau=-2^{l_i},\\
   0, & \textnormal{otherwise.}
      \end{cases}
 \end{split}
\end{equation}
From (\ref{L_1=T+T_i}), (\ref{Tderi}) and (\ref{T_iderivation}), we have
\begin{equation}\label{L_1derivation}
\begin{split}
\mathcal{L}_1&=T+\sum_{i=1}^pT_i\\
&=
 \begin{cases}
  2^{m+1}\displaystyle\sum_{i=1}^pN_i+2^{m+1}M, & \tau=0,\\
  \omega_q^{g_{l_i}}2^{m}\displaystyle\sum_{\textbf{c}\in S_{N_i}}\omega_q^{L^{l_i}_{\textbf{c}}}, & \tau\!=\! 2^{l_i},i\!\!=\!\!1,2,\hdots,p,\\
   \omega_q^{-g_{l_i}}2^{m}\displaystyle\sum_{\textbf{c}\in S_{N_i}}\omega_q^{-L^{l_i}_{\textbf{c}}}, & \tau\!=\!-2^{l_i},i\!\!=\!\!1,2,\hdots,p,\\
   0, & \textnormal{otherwise.}
 \end{cases}
 \end{split}
\end{equation}
To find $\mathcal{L}_2$, we start with
\begin{equation}\label{L_2derived}
\begin{split}
&\displaystyle \sum_{\textbf{d}}\displaystyle C \left( \left(f+\frac{q}{2}(\mathbf{d}\cdot\mathbf{x}_J+d''e_2)\right)\big\arrowvert_{\textbf{x}_J=\textbf{c}_1} ,\right.\\&\left.~~~~~~~~~~~~~~~~~~~~
 \left(f+\frac{q}{2}(\mathbf{d}\cdot\mathbf{x}_J+d''e_2)\right)\big\arrowvert_{\textbf{x}_J=\textbf{c}_2} \right) (\tau)\qquad\\
&=\displaystyle\sum_{\textbf{d}}(-1)^{\textbf{d}\cdot(\textbf{c}_1+\textbf{c}_2)}C \left( \left(f+\frac{q}{2}(d''e_2)\right)\big\arrowvert_{\textbf{x}_J=\textbf{c}_1} ,\right.\\&\left.~~~~~~~~~~~~~~~~~~~~~~~~~~~~~~~~~~
 \left(f+\frac{q}{2}(d''e_2)\right)\big\arrowvert_{\textbf{x}_J=\textbf{c}_2} \right) (\tau)\\
&= C \left( \left(f+\frac{q}{2}(d''e_2)\right)\big\arrowvert_{\textbf{x}_J=\textbf{c}_1} ,\right. \\& \left.~~~~~~~~~
 \left(f+\frac{q}{2}(d''e_2)\right)\big\arrowvert_{\textbf{x}_J=\textbf{c}_2} \right) (\tau)
 \displaystyle\sum_{\textbf{d}}(-1)^{\textbf{d}\cdot(\textbf{c}_1+\textbf{c}_2)}\\
&=0\!\!\quad\!\!\forall \tau.
\end{split}
\end{equation}
Therefore, from (\ref{L_2}) and (\ref{L_2derived}), we have
\begin{equation}\label{L_2derivation}
\begin{split}
\mathcal{L}_2& =\displaystyle \sum_{\textbf{d}d''}\displaystyle \sum_{\textbf{c}_1\neq \textbf{c}_2}C \left( \left(f+\frac{q}{2}(\mathbf{d}\cdot\mathbf{x}_J+d''e_2\right)\big\arrowvert_{\textbf{x}_J=\textbf{c}_1}
,\right.\\&~~~~~~~~~~~~~~~~~ \left. \left(f+\frac{q}{2}(\mathbf{d}\cdot\mathbf{x}_J+d''e_2)\right)\big\arrowvert_{\textbf{x}_J=\textbf{c}_2} \right) (\tau)\\
&=0,~\forall \tau.
\end{split}
\end{equation}
By substituting (\ref{L_1derivation}) and (\ref{L_2derivation}) into (\ref{A=L_1+L_2}), we complete the proof.
\section{Enumeration of complementary sequences with maximum PMEPR $4$, $6$, and $8$}
In this section, we have given the derivarions on enumeration of complementary sequences with maximum PMEPR $4$, $6$, and $8$.
\subsection{Enumeration of complementary sequences with maximum PMEPR $4$ by \textit{Corollary 1}}
Let $\pi$ be a permutation of the set $\mathcal{S}_{\alpha,l}=\{0,1,\hdots,m-1\}\setminus\{\alpha,l\}$, where $\alpha,l\in\{0,1,\hdots,m-1\}$,
and $\alpha\neq l$. We define a quadratic GBF $Q_\pi$ as follows:
\begin{equation}\label{gbfc1}
 Q_\pi=\frac{q}{2}\displaystyle\sum_{i=0}^{m-4}x_{\pi(i)}x_{\pi(i+1)}.
\end{equation}
Therefore, $Q_\pi$ is a quadratic GBF over the variable $\{x_0,x_1,\hdots,x_{m-1}\}\setminus\{x_\alpha,x_l\}$.
There exist $\frac{(m-2)!}{2}$ permutations for which we have $\frac{(m-2)!}{2}$ distinct quadratic GBFs as given in (\ref{gbfc1}).
Let $\pi_1,\pi_2,\hdots,\pi_{\frac{(m-2)!}{2}}$ be the $\frac{(m-2)!}{2}$ distinct permutations and $Q_{\pi_1},Q_{\pi_2},\hdots,Q_{\pi_{\frac{(m-2)!}{2}}}$,
the corresponding GBFs. Now, we define a GBF $f:\{0,1\}^m\rightarrow \mathbb{Z}_q$ as follows:
\begin{equation}\label{gbfc2}
\begin{split}
 f=x_\alpha Q_{\pi_u}+(1-x_\alpha)Q_{\pi_v}+\sum_{\beta=0}^{m-3}a_{\alpha,\pi_u(\beta)}x_\alpha x_{\pi_u(\beta)}\\+\frac{q}{2}x_\alpha x_l
 +\sum_{i=0}^{m-1}g_ix_i+g',
 \end{split}
\end{equation}
where $u,v\in\left\{1,2,\hdots,\frac{(m-2)!}{2}\right\}$, $u\neq  v$, $a_{\alpha,\pi_u(\beta)}\in\mathbb{Z}_q$, $g_i\in\mathbb{Z}_q$, and
$g'\in\mathbb{Z}_q$. For a fixed choice of $\alpha,l,u,v$ and in order to avoid repetations of the same GBFs, we consider $a_{\alpha,\pi_u(\beta)}\in\mathbb{Z}_q$
for $\beta\in\{1,2,\hdots,m-4\}$, and $a_{\alpha,\pi_u(\beta)}\in\mathbb{Z}_q\setminus\{\frac{q}{2}\}$ for $\beta\in\{0,m-3\}$. For a fixed
choice of $\alpha$, $l$ and by varying $u,v$, we have $\frac{(m-2)!}{2}\left[\frac{(m-2)!}{2}-1\right]$ distinct GBFs in the form
$x_\alpha Q_{\pi_u}+(1-x_\alpha)Q_{\pi_v}$. Therefore, for fixed $\alpha$ and $l$, we get at least
$\frac{(m-2)!}{2}\left[\frac{(m-2)!}{2}-1\right]q^{m-4}(q-1)^2q^{m+1}=\frac{(m-2)!}{2}\left[\frac{(m-2)!}{2}-1\right]q^{2m-3}(q-1)^2$ distinct GBFs.
It is noted that $\alpha$ can be selected in  $m$ ways and for each choice of
$\alpha$, $l$ can be selected in $m-1$ ways. Therefore, there exist at least $m(m-1)\frac{(m-2)!}{2}\left[\frac{(m-2)!}{2}-1\right]q^{2m-3}(q-1)^2=
\frac{m!}{2}\left[\frac{(m-2)!}{2}-1\right]q^{2m-3}(q-1)^2$ distinct GBFs.

From (\ref{gbfc2}), it is clear that either $G(f\arrowvert_{x_\alpha=0})$ or $G(f\arrowvert_{x_\alpha=1})$ contains a path over
the vertices $\{x_0,x_1,\hdots,$ $x_{m-1}\}\setminus\{x_\alpha,x_l\}$ and one isolated vertex $x_l$. The paths in $G(f\arrowvert_{x_\alpha=0})$ and
$G(f\arrowvert_{x_\alpha=1})$ are identified by $G(Q_v)$ and $G(Q_u)$, respectively. From (\ref{gbfc2}),
$L_{x_\alpha}^l=\frac{q}{2}x_\alpha$ which gives $L_0^l=0$ and $L_1^l=\frac{q}{2}$. Hence, $f$ satisfies the properties given in \textit{Corollary 1}
for $k=1$. Therefore, we obtain $\frac{m!}{2}\left[\frac{(m-2)!}{2}-1\right]q^{2m-3}(q-1)^2$ distinct GBFs of
order three whose corresponding sequences have PMEPRs upper bounded by 4.

\subsection{Enumeration of complementary sequences with maximum PMEPR $8$ by \textit{Corollary 1}}
Let $\pi'$ be a permutation of the set $\mathcal{S}_{\alpha_1,\alpha_2,l}=\{0,1,\hdots,m-1\}\setminus\{\alpha_1,\alpha_2,l\}$, where
$\alpha_1,\alpha_2$, and $l\in\{0,1,\hdots,m-1\}$ are distinct. We define a quadratic GBF $Q_{\pi'}$ as follows:
 \begin{equation}\label{gbfc3}
 Q_{\pi'}=\frac{q}{2}\displaystyle\sum_{i=0}^{m-5}x_{\pi'(i)}x_{\pi'(i+1)}.
\end{equation}
There exist $\frac{(m-3)!}{2}$ permutations for which we have $\frac{(m-3)!}{2}$ distinct quadratic GBFs of the form given in (\ref{gbfc3}).
Let $\pi'_1,\pi'_2,\hdots,\pi'_{\frac{(m-3)!}{2}}$ be the permutations and $Q_{\pi'_1},Q_{\pi'_2},\hdots,Q_{\pi'_{\frac{(m-3)!}{2}}}$,
the corresponding GBFs. We define the GBF $f':\{0,1\}^m\rightarrow \mathbb{Z}_q$ as follows:
\begin{equation}\label{gbfc4}
\begin{split}
 f'=&\left(x_{\alpha_1}x_{\alpha_2}\!\!+\!\!(1\!-\!x_{\alpha_1})(1\!-\!x_{\alpha_2})\right) Q_{\pi'_{u_1}}\!\!+\!\!\left(x_{\alpha_1}(1-x_{\alpha_2})\right.\\&\left.+x_{\alpha_2}(1-x_{\alpha_1})\right)Q_{\pi'_{v_1}}\!\!+\!\!\sum_{\beta=0}^{m-4}a'_{\alpha_1,\pi'_{u_1}(\beta)}x_{\alpha_1} x_{\pi'_{u_1}(\beta)}
 \\&
 +\sum_{\beta=0}^{m-4}a''_{\alpha_2,\pi'_{v_1}(\beta)}x_{\alpha_2} x_{\pi'_{v_1}(\beta)}+bx_{\alpha_1}x_{\alpha_2}+L^l_{\textbf{x}_J}x_l\\&+\sum_{i=0}^{m-1}g_ix_i+g',
 \end{split}
\end{equation}
where $u_1,v_1\in\left\{1,2,\hdots,\frac{(m-3)!}{2}\right\}$, $u_1\neq  v_1$, $a'_{\alpha_1,\pi'_{u_1}(\beta)}\in\mathbb{Z}_q$, $a''_{\alpha_2,\pi'_{v_1}(\beta)}\in\mathbb{Z}_q$, $b\in\mathbb{Z}_q$, $g_i\in\mathbb{Z}_q$,
$g'\in\mathbb{Z}_q$, and $\textbf{x}_J=(x_{\alpha_1},x_{\alpha_2})\in\{0,1\}^2$. The term $L^l_{\textbf{x}_J}$ present in (\ref{gbfc4})
can be selected in $3$ ways which are $\frac{q}{2}x_{\alpha_1}$,
$\frac{q}{2}x_{\alpha_2}$, and $\frac{q}{2}(x_{\alpha_1}+x_{\alpha_2})$. For a fixed choice of $\alpha_1,\alpha_2,l,u_1,v_1$ and
to avoid repetations of the same GBFs, we consider $a'_{\alpha_1,\pi'_{u_1}(\beta)},a''_{\alpha_2,\pi'_{v_1}(\beta)}\in\mathbb{Z}_q$
for $\beta\in\{1,2,\hdots,m-5\}$, $a'_{\alpha_1,\pi'_{u_1}(\beta)}\in\mathbb{Z}_q\setminus\{\frac{q}{2}\}$ for $\beta\in\{0,m-4\}$. We fixed
$a''_{\alpha_2,\pi'_{v_1}(0)}$ and $a''_{\alpha_2,\pi'_{v_1}(m-4)}$ in $\mathbb{Z}_q\setminus\{0,\frac{q}{2}\}$.
For a fixed choice of $\alpha_1$, $\alpha_2$, $l$ and by varying $u_1,v_1$, we obtain $\frac{(m-3)!}{2}\left[\frac{(m-3)!}{2}-1\right]$ distinct GBFs in the form
$\left(x_{\alpha_1}x_{\beta_1}+(1-x_{\alpha_1})(1-x_{\beta_1})\right) Q_{\pi'_{u_1}}+\left(x_{\alpha_1}(1-x_{\beta_1})+x_{\beta_1}(1-x_{\alpha_1})\right)Q_{\pi'_{v_1}}$.

From (\ref{gbfc4}), we obtain
at least $\frac{3m!}{4}\left[\frac{(m-3)!}{2}-1\right]q^{3m-8}(q-1)^2$ distinct GBFs.
It is clear that each of $G\left(f\arrowvert_{(x_{\alpha_1}, x_{\alpha_2})=(0,0)}\right)$, $G\left(f\arrowvert_{(x_{\alpha_1}, x_{\alpha_2})=(0,1)}\right)$,
$G\!\left(f\arrowvert_{(x_{\alpha_1}, x_{\alpha_2})=(1,0)}\right)$, and $G\!\left(f\arrowvert_{(x_{\alpha_1}, x_{\alpha_2})=(0,0)}\right)$ contains
a path over $m-3$ vertices and one isolated vertex $x_l$. The paths in $G\!\left(\!f\arrowvert_{(x_{\alpha_1}, x_{\alpha_2})=(0,0)}\!\right)$ and
$G\!\left(\!f\arrowvert_{(x_{\alpha_1}, x_{\alpha_2})=(1,1)}\!\right)$ are identified by $G(Q_{\pi'_{u_1}})$, while the paths in
$G\left(f\arrowvert_{(x_{\alpha_1}, x_{\alpha_2})=(0,1)}\right)$ and
$G\left(f\arrowvert_{(x_{\alpha_1}, x_{\alpha_2})=(1,0)}\right)$ are identified by $G(Q_{\pi'_{v_1}})$.
For $L^l_{\textbf{x}_J}=\frac{q}{2}x_{\alpha_1}$,
$L^l_{\textbf{x}_J}$ equals $0$ when $\textbf{x}_J\in \{(0,0),(0,1)\}$ and $L^l_{\textbf{x}_J}$ equals $\frac{q}{2}$ when $\textbf{x}_J\in \{(1,0),(1,1)\}$.
For the remaining two
choices of $L^l_{\textbf{x}_J}$, we can verify that
there exist exactly two vectors in $\{0,1\}^2$ for which $L^l_{\textbf{x}_J}\equiv 0~(\!\!\!\!\mod q)$ and $L^l_{\textbf{x}_J}\equiv \frac{q}{2}~(\!\!\!\!\mod q)$ for
another two vectors in $\{0,1\}^2$. Therefore, the GBF $f$, given in (\ref{gbfc4}), satisfies all the properties specified in
\textit{Corollary 1} for
$k=2$, and $p=1$. Hence, we have at least $\frac{3m!}{4}\left[\frac{(m-3)!}{2}-1\right]q^{3m-8}(q-1)^2$ complementary sequences with the PMEPR
upper bounded by $8$. Following \textit{Corollary 1}, more GBFs and corresponding complementary sequences may be constructed specially by
taking $k=2$, and $p=2$.
To compare our proposed code-rate with [4], we consider only $\frac{3m!}{4}\left[\frac{(m-3)!}{2}-1\right]q^{3m-8}(q-1)^2$ complementary sequences of PMEPR at most $8$ by
\textit{Corollary 1}.
\subsection{Enumeration of complementary sequences with maximum PMEPR $6$ by \textit{Corollary 2}}
In the Subsection \textit{A} of this section, we have defined $\mathcal{S}_{\alpha,l}$, $\pi$, $Q_{\pi_u}$, where
$u\in\left\{1,2,\hdots,\frac{(m-2)!}{2}\right\}$, which
will be used to count complementary sequences with maximum PMEPR $6$.

Let $\pi''$ be a permutation of the set $\mathcal{S}'_{\alpha}=\{0,1,\hdots,m-1\}\setminus\{\alpha\}$. We define a quadratic GBF $Q_{\pi''}$ as follows:
\begin{equation}\label{gbfc5}
 Q_{\pi''}=\frac{q}{2}\displaystyle\sum_{i=0}^{m-3}x_{\pi''(i)}x_{\pi''(i+1)}.
\end{equation}
Let $\pi''_1,\pi''_2,\hdots,\pi''_{\frac{(m-1)!}{2}}$ be the permutations and $Q_{\pi''_1},Q_{\pi''_2},\hdots,Q_{\pi''_{\frac{(m-1)!}{2}}}$,
the corresponding GBFs. We define the GBF $f'':\{0,1\}^m\rightarrow \mathbb{Z}_q$ as follows:
\begin{equation}\label{gbfc6}
\begin{split}
 f''=x_\alpha Q_{\pi_u}+(1-x_\alpha)Q_{\pi''_{v'}}+\sum_{\beta=0}^{m-2}b_{\alpha,\pi''_{v'}(\beta)}x_\alpha x_{\pi''_{v'}(\beta)}
 \\+\sum_{i=0}^{m-1}g_ix_i+g',
 \end{split}
\end{equation}
where $u\in\left\{1,2,\hdots,\frac{(m-2)!}{2}\right\}$, $v'\in\left\{1,2,\hdots,\frac{(m-1)!}{2}\right\}$, $b_{\alpha,\pi''_{v'}}\in \mathbb{Z}_q$, $g_i\in\mathbb{Z}_q$,
and $g'\in\mathbb{Z}_q$. For a fixed choice of $\alpha,l,u,v'$ and to avoid repetations of the same GBFs, we consider $b_{\alpha,\pi''_{v'}(\beta)}\in\mathbb{Z}_q$
for $\beta\in\{1,2,\hdots,m-3\}$, and $b_{\alpha,\pi''_{v'}(\beta)}\in\mathbb{Z}_q\setminus\{\frac{q}{2}\}$ for $\beta\in\{0,m-2\}$.

From (\ref{gbfc6}),
we obtain at least $\left[\frac{m!(m-2)!(m-1)}{4}\right]q^{2m-2}(q-1)^2$ distinct GBFs.
It is clear that $G(f''\arrowvert_{x_\alpha=0})$ is a path identified by $G(Q_{\pi''_{v'}})$,
$G(f''\arrowvert_{x_\alpha=1})$ contains a path and one isolated vertex $x_l$. The path in $G(f''\arrowvert_{x_\alpha=1})$ is identified
by $G(Q_{\pi_u})$. Therefore, the GBF $f''$, given in (\ref{gbfc6}), satisfies all the properties specified in
\textit{Corollary 2} for $k=1$ and $p=1$.
Hence, we obtain at least $\left[\frac{m!(m-2)!(m-1)}{4}\right]q^{2m-2}(q-1)^2$ complementary sequences with the PMEPR
upper bounded by $6$.
\subsection{Enumeration of complementary sequences with maximum PMEPR $8$ by \textit{Corollary 2}}
Let $\pi^{l_1}$ be a permutaion of $\mathcal{S}_{\alpha,l_1}$ and $\pi^{l_2}$ be a permutaion of $\mathcal{S}_{\alpha,l_2}$, where $\alpha$, $l_1$,
and $l_2$ are three distinct integer values from $\{0,1,\hdots,m-1\}$. We define the quadratic GBFs $Q_{\pi^{l_1}}$ and $Q_{\pi^{l_2}}$
as follows:
\begin{equation}\label{gbfc7}
\begin{split}
 Q_{\pi^{l_1}}&=\frac{q}{2}\displaystyle\sum_{i=0}^{m-4}x_{\pi^{l_1}(i)}x_{\pi^{l_1}(i+1)},\\
 Q_{\pi^{l_2}}&=\frac{q}{2}\displaystyle\sum_{i=0}^{m-4}x_{\pi^{l_2}(i)}x_{\pi^{l_2}(i+1)}.
 \end{split}
\end{equation}
Let $Q_{\pi^{l_1}_1},Q_{\pi^{l_1}_2},\hdots,Q_{\pi^{l_1}_{\frac{(m-2)!}{2}}}$ be the quadratic GBFs corresponding to $\pi^{l_1}_1,\pi^{l_1}_2,\hdots,\pi^{l_1}_{\frac{(m-2)!}{2}}$
respectively, and $Q_{\pi^{l_2}_1},Q_{\pi^{l_2}_2},\hdots,Q_{\pi^{l_2}_{\frac{(m-2)!}{2}}}$ be the quadratic GBFs corresponding to
$\pi^{l_2}_1,\pi^{l_2}_2,\hdots,\pi^{l_2}_{\frac{(m-2)!}{2}}$ respectively.
Let us define a GBF $f''':\{0,1\}^m\rightarrow \mathbb{Z}_q$ as follows:
\begin{equation}\label{gbfc8}
\begin{split}
 f'''=x_\alpha Q_{\pi^{l_1}_u}+(1-x_\alpha)Q_{\pi^{l_2}_v}+\sum_{\beta=0}^{m-3}b'_{\alpha,\pi^{l_2}_v(\beta)}x_\alpha x_{\pi^{l_2}_v(\beta)},
 \\+\sum_{i=0}^{m-1}g_ix_i+g',
 \end{split}
\end{equation}
where $u,v\in\left\{1,2,\hdots,\frac{(m-2)!}{2}\right\}$, $b'_{\alpha,\pi^{l_2}_v(\beta)}\in \mathbb{Z}_q$ for $\beta=1,2,\hdots,m-4$,
$b'_{\alpha,\pi^{l_2}_v(\beta)}\in \mathbb{Z}_q\setminus\{\frac{q}{2}\}$ for $\beta=0,m-3$, $g_i\in\mathbb{Z}_q$, and $g'\in\mathbb{Z}_q$.
Note that $\alpha$ can be selected in $m$ ways and for each choice of $\alpha$, $l_1$ can be selected in $m-1$ ways. In order to avoid repetations of
the same GBF, we choose $l_1$ in one way. Therefore, for each choice of $\alpha$ and for the fixed choice of $l_1$, $l_2$ can be chosen in
$m-2$ ways. From (\ref{gbfc8}), we obtain at least $\left[\frac{(m-2)!}{2}\right]^2 q^{2m-3}(q-1)^2$ distinct GBFs.

$G(f'''\arrowvert_{x_\alpha=0})$ contains a path identified by $G(Q_{\pi^{l_2}_v})$ and one isolated vertex $x_{l_1}$.
Also, $G(f'''\arrowvert_{x_\alpha=1})$ contains a path identified by $G(Q_{\pi^{l_1}_v})$ and one isolated vertex $x_{l_2}$.
Therefore, the GBF $f'''$ satisfies all the properties given in \textit{Corollary 2} for $k=1$ and $p=2$. Hence, we obtain at least
$\left[\frac{(m-2)!}{2}\right]^2 q^{2m-3}(q-1)^2$ complementary sequences with the PMEPR upper bounded by $8$.
\end{appendices}
\bibliographystyle{IEEEtran}
\bibliography{Palash1}
\end{document}